\documentclass[fleqn,10pt]{wlscirep}
\usepackage[utf8]{inputenc}
\usepackage[T1]{fontenc}
\usepackage{amsmath}
\usepackage{amssymb}
\usepackage{graphicx}
\usepackage{braket}
\usepackage{svg}
\usepackage{xfrac}  
\usepackage{subcaption}
\usepackage{authblk}
\usepackage{algorithm2e}
\usepackage{float}
\RestyleAlgo{ruled}
\SetKwComment{Comment}{/* }{ */}
\SetKwInOut{Parameter}{parameter}

\title{Hybrid adiabatic quantum computing for tomographic image reconstruction - opportunities and limitations}

\author[1,2]{Merlin A. Nau}
\author[3]{A. Hans Vija}
\author[3]{Wesley Gohn}
\author[1,2]{Maximilian P. Reymann}
\author[1]{Andreas K. Maier}
\affil[1]{Friedrich-Alexander Universit\"at Erlangen-N\"urnberg, Pattern Recognition Lab, Erlangen, Germany}
\affil[2]{Siemens Healthineers GmbH, Forchheim, Germany}
\affil[3]{Siemens Medical Solutions USA, Inc, Hoffman Estates, USA}

\affil[*]{merlin.nau@fau.de}

\begin{abstract}
Our goal is to reconstruct tomographic images with few measurements and a low signal-to-noise ratio. In clinical imaging, this helps to improve patient comfort and reduce radiation exposure. As quantum computing advances, we propose to use an adiabatic quantum computer and associated hybrid methods to solve the reconstruction problem. Tomographic reconstruction is an ill-posed inverse problem. We test our reconstruction technique for image size, noise content, and underdetermination of the measured projection data. We then present the reconstructed binary and integer-valued images of up to $32 \times 32$ pixels.
The demonstrated method competes with traditional reconstruction algorithms and is superior in terms of robustness to noise and reconstructions from few projections. We postulate that hybrid quantum computing will soon reach maturity for real applications in tomographic reconstruction. Finally, we point out the current limitations regarding the problem size and interpretability of the algorithm.
\end{abstract}
\begin{document}

\flushbottom
\maketitle
%
%
\thispagestyle{empty}

\section*{Introduction}

Quantum computing (QC) presents a new algorithmic paradigm that has gained much attention over the past decades. Although practical applications are minimal, QC offers enormous potential for complex computations and runtime speed up~\cite{nielsen2002quantum}. Quantum supremacy has been claimed in 2019 and 2021, respectively~\cite{arute2019quantum, wu2021strong}. 

One common QC model is the gate-base model~\cite{nielsen2002quantum}. However, not all quantum computers build upon this model. Different physical implementations can utilize the same quantum mechanics needed to realize QC. Other QC principles include measurement-based, adiabatic~\cite{aharonov2008adiabatic}, and topological QC.

We are utilizing adiabatic QC produced by D-Wave. However, current implementations violate adiabatic rules, so D-Wave's quantum computer is not universal. Due to the current limitation of the quantum computer, we utilize D-Wave's corresponding hybrid solvers, which combine both quantum and classical computing power~\cite{raymond2022hybrid}. 

In this work, we investigate the task of QC-based tomographic image reconstruction. Tomographic imaging is the process of imaging sections hidden within an object~\cite{maier2018medical}. Consequently, tomographic imaging techniques are fundamental in numerous fields, from radiology and materials science to astrophysics. In particular, we are interested in emission tomography (ET), which has a very low signal-to-noise ratio~\cite{wernick2004emission}. We also point out significant differences in transmission tomography (TT). For simplicity, we generally refer to the term tomography. 

The primary problem setting in tomographic imaging is reconstructing the inner sections. The reconstruction of the underlying object is an inverse problem, which is usually ill-posed and can degrade in terms of completeness or noise~\cite{wernick2004emission}. In order to reconstruct ET images at today's scale of $512 \times 512$, advancements in classical computing hardware were necessary. Similarly, we expect QC to boost performance in tomographic reconstruction as the hardware matures.

In the following sections, we will give an overview of QC, particularly for the D-Wave quantum computer we utilize for our applications. Further, we will provide the reader with emission and transmission tomographic imaging fundamentals. More to the point, we will describe the inverse problem in tomographic reconstruction and how state-of-the-art methods solve it today. After that, we elaborate on our quantum annealing (QA)-based reconstruction technique and present binary and integer value reconstruction results.

\section*{Quantum Computing}

\begin{figure}
    \centering
    \includegraphics[scale=0.6]{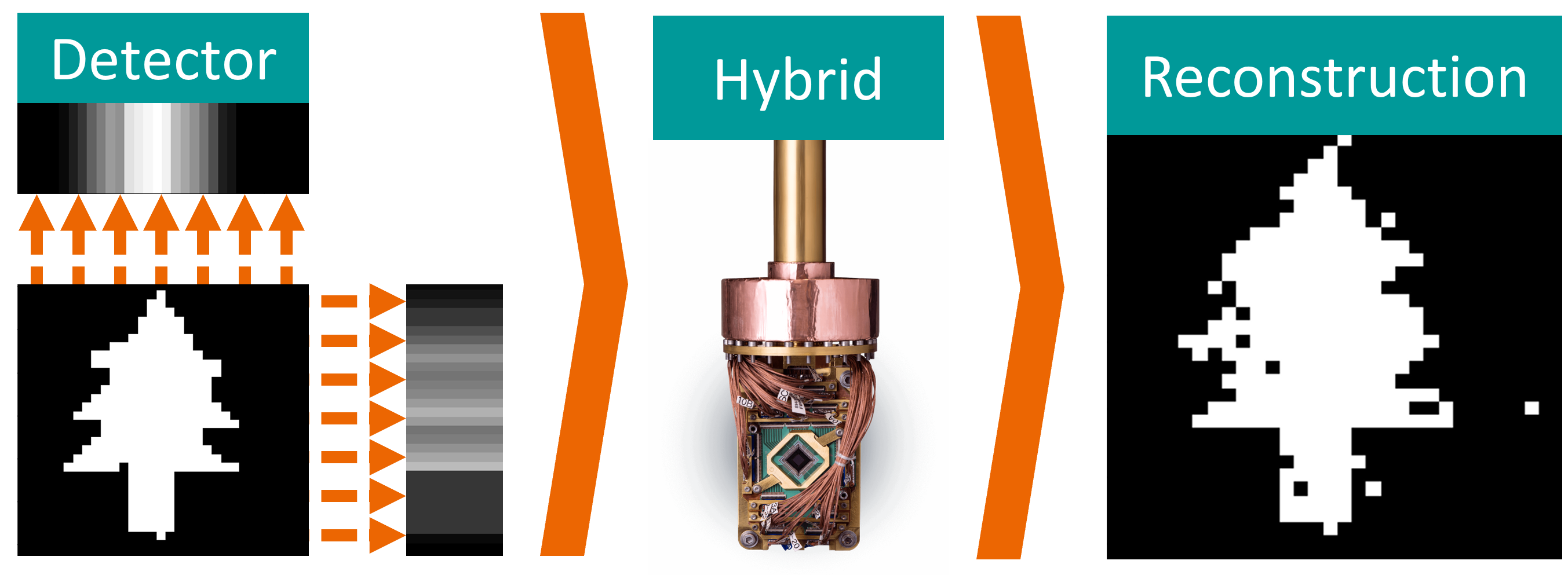}
    \caption{Simulation and reconstruction of a two-view binary tomographic problem using hybrid quantum annealing.}
    \label{fig:radon}
\end{figure}

QC is a new and innovative computing paradigm. Instead of using classical electronic bits, quantum computers utilize quantum bits (qubits) to exploit the quantum mechanical principles of superposition, entanglement and interference. Using these paradigms, a quantum computer with $N$ qubits can be in $2^N$ states simultaneously, compared to one state for a classical computer~\cite{rieffel2000introduction}. This advantage ultimately leads to a benefit in terms of run time speed-up while enabling computations that are impossible on a classical computer. The concept of universal QC is accomplishable by several models. The total number of qubits is currently limited to 5640 qubits on the QA-based D-Wave Advantage2 system. For gate-based systems the current record is set by the IBM eagle system with 127 superconducting qubits~\cite{chow2021ibm}. 

\subsection*{Adiabatic Quantum Computing}

The fundamental assumption of AQC is that a physical system constantly evolves to its lowest energy over time~\cite{albash2018adiabatic, arute2019quantum} - associated with the global minimum of the optimization landscape. In contrast to gate-based QC, AQC does not perform unitary operations through gates on single or multiple qubits~\cite{van2001powerful}. Instead, we map the problem to the quantum computer with a problem-specific Hamiltonian~\cite{farhi2001quantum}. The Hamiltonian describes an energy spectrum of the system and the set of viable solutions. In theory, AQC is still equivalent to gate-based QC and, therefore, universal~\cite{aharonov2008adiabatic}. 

The adiabatic theorem, states that if we initialize a quantum system with a ground state  $\hat{H_0}$ and let it evolve with time $t$ for a fixed duration $T$, we will end up in the ground state $\hat{H_1}$, which is associated with the lowest energy solution~\cite{born1928beweis}: 

\begin{equation}
\label{eqn:adiabatic}
     \hat H (t) = (1 - \sfrac{t}{T}) \hat{H_0} + \sfrac{t}{T} \hat{H_1}
\end{equation}
The main limitation of AQC is the $\Delta$-gap~\cite{jansen2007bounds}. The $\Delta$-gap refers to the minimum spectral gap of the problems Hamiltonian, which is the difference between the lowest and second-lowest energy levels. The speed limit $T$ is then calculated as:  

\begin{equation}
\label{eqn:timelimit}
     T = \frac{1}{min(\Delta (t))^2}
\end{equation}

\subsection*{Quantum Annealing}
QA is the current realization of AQC. The system is initialized in a superposition state. Subsequently, the problem formulation is embedded in the hardware such that the system's ground state is the solution to the problem~\cite{finnila1994quantum}. But, how does QA overcome the $\Delta$-gap?

 In short, it does not. Physical realizations of AQC usually let the system evolve multiple times for a specified time~\cite{mcgeoch2014adiabatic}. After initialization, the system repeatedly anneals for a specified annealing time $t_a$. This way, a sample set is formed containing the samples, the associated energy level, and the number of times the solution occurred~\cite{koshka2020comparison}. 
 
\begin{figure}
\centering
\includegraphics[scale=0.65]{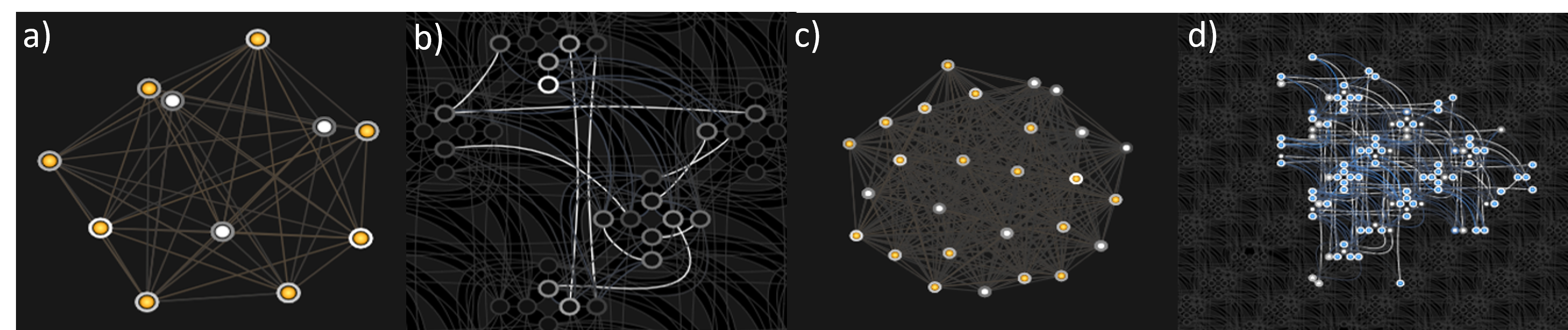}  
\caption{Graphs and physical embeddings on the QPU for binary reconstruction problems. a) and c) show the directed graph for binary tomographic problems for image sizes $4 \times 4$ and $8 \times 8$, respectively. b) and d) show the embedding on the QPU topology for a) and c), respectively.}
\label{fig:embedding}
\end{figure}

QA builds upon the Hamiltonian of the Ising model: 

\begin{equation}
\label{eqn:ising}
    E_{Ising}(s) = \sum_{i=1}^{N} h_i s_i + \sum_{i=1}^{N} \sum_{i \neq j}^{N}   J_{i,j} s_i s_j  
\end{equation}
The variables $s_i$ of the Ising model are either spin-up or spin-down ${-1, 1}$, which resemble the eigenvalues of Pauli matrices~\cite{nielsen2002quantum}. Two variables $s_i$ and $s_j$ can have quadratic interactions $J_{i,j}$, known as the coupling strength. Further, one variable can have a linear bias $h_i$. Every Ising model is translatable to a Quadratic Unconstrained Binary Optimization (QUBO) problem, with variables $x_i$ being binary {0, 1}, and vice-versa. QUBO problems can be NP-hard and are hard to solve using classical computers~\cite{lucas2014ising}. We can describe a QUBO using an $N \times N$ upper-triangular matrix with the bias term on its diagonal and the quadratic interaction as the upper-triangular values: 

\begin{equation}
\label{eqn:qubo}
    f(x) = \sum_{i} Q_{i,j} x_{i} + \sum_{i<j} Q_{i,j} x_i x_j  
\end{equation}
The goal is to minimize the QUBO's objective. In matrix notation, this results in the following:

\begin{equation}
\label{eqn:qubomatrix}
    \min_{x \in \{ 0,1 \}^n} x^T Q x
\end{equation}
The connectivity of the qubits on the annealer's topology limits the interaction between qubits, see Fig.~\ref{fig:embedding}. D-Wave has proposed different topologies over the last years, such as the Chimaera~\cite{boothby2016fast, king2018observation} or Pegasus graph. By using D-Wave's Ocean interface~\cite{dwave2022ocean}, we can embed and run the problem on the quantum annealer using the Leap cloud service~\cite{dwave2022leap}.  

\section*{Tomographic Imaging}

Tomographic reconstruction is a multidimensional inverse problem. The problem is estimating an object inside by only acquiring non-invasive measurements. The tomographic imaging process is similar to any linear digital imaging system, which maps a continuous domain to a discrete domain~\cite{barrett2013foundations}. However, in practice, the image formation process is defined in a discrete-to-discrete forward model:

\begin{equation}
\label{eqn:forwardproblem}
    y = M x
\end{equation}
Here $y$ represents our measurement, $M$ is the system matrix describing the action of the linear imaging system, and $x$ represents the imaged object. In the realistic case of an imperfect imaging system and setting, a common representation includes additive noise~\cite{barrett2013foundations}:

\begin{equation}
\label{eqn:forwardproblemnoise}
    y = M x + n
\end{equation}
Note that not every noise is additive, as the noise can also be signal-dependent on $x$. To evaluate the stability of reconstruction, one can compare reconstructions from two measurements, $y_1$ and $y_2$, which are different in terms of their noise realizations. If the noise realizations are similar, the stability should estimate how close the reconstructed images $\hat x_1$ and $\hat x_2$ are. Additionally, $\alpha$ is a constant, accepting a certain tolerance.  Stability defines as: 

\begin{equation}
\label{eqn:stability}
    \lVert \hat x_1 - \hat x_2 \rVert_2 \leq \alpha \lVert y_1 - y_2 \rVert_2
\end{equation}
This paper focuses on radiation-based tomography, where the foundation is the Radon transform~\cite{cormack1963representation, beylkin1987discrete}. However, the forward and inverse problem is similar to, e.g., magnetic resonance imaging (MRI). Our method is applicable to any imaging modality with a \textbf{matrix-based-forward model}. In radiation tomography, one distinguishes between ET and TT. 
\newline
\textbf{Emission tomography.} In ET, the object of interest emits radiation from the inside. 
In clinical imaging, one typically injects a patient with a radioactive tracer. The tracer carries a radioactive isotope to a target location, where it emits radiation. Instead of visualizing the anatomy, ET reveals the metabolic and biochemical function of the underlying tissue~\cite{wernick2004emission}.
\newline
\textbf{Transmission tomography.} In transmission tomography, or Computed Tomography (CT), the projection views are captured by placing an X-ray source on one side of the object and an X-ray detector on the other side~\cite{hsieh2003computed}~\cite{maier2018medical}. The X-rays emitted from the source are attenuated by the matter and captured by the detector.\newline
\textbf{Discrete tomography.} Discrete tomography is a specialized case of tomography, where the image $x$ consists of binary pixels $x_i \in {0, 1}$~\cite{herman2012discrete}. This tomography case can apply to homogeneous scanning materials for non-destructive testing~\cite{herman2012discrete}. Because the complexity of the optimization is constrained to binary variables, the task's difficulty is to reconstruct with as few views as possible.
\newline
\textbf{Tomographic Reconstruction.} In this section, we outline different algorithms to invert a forward problem to find the original image $x$. The system matrix $M$ is defined by the imaging system and can model physical effects or prior knowledge about imperfections. It describes the conditional probability that  $y_j$ detects an emission of $x_i$ for each matrix entry:

\begin{equation}
\label{eqn:reconstructionmatrixexample}
M =  p(\text{detected in y } | \text{ emitted in x})
\end{equation}
Thus, in image reconstruction, we are interested in finding the inverse of M to obtain the original image:
\begin{equation}
\label{eqn:inverseproblem}
    x = M^{-1}y
\end{equation}

In practice, the system matrix $M$ is sparse, exceptionally large, singular, ill-posed, and non-square. Moreover, the projection images suffer from various sources of noise, most prominently Poisson noise. One conventional method to reconstruct tomographic images is Filtered BackProjection (FBP), an analytical and linear approach~\cite{pan2009commercial}. Other approaches to reconstructing tomographic images are iterative reconstruction algorithms~\cite{bruyant2002analytic}. In contrast to FBP, these algorithms are non-linear and seek to minimize the projection difference by repetitively applying back projections, updates, and forward projections~\cite{lange1984reconstruction}. Three examples of iterative reconstruction techniques include Maximum Likelihood Expectation Maximization (ML-EM)~\cite{dempster1977maximum, lange1984reconstruction}, Conjugate Gradient (CG)~\cite{hestenes1952methods}, and Simultaneous Algebraic Reconstruction Technique (SART)~\cite{andersen1984simultaneous}. SART is an algebraic iterative reconstruction algorithm that performs additive and iterative updates from single projections. As the image size for the QA-based reconstruction is inherently limited by the size of the annealer, we also consider the Moore-Penrose general pseudoinverse (PI) as a reconstruction technique. The Moore-Penrose is defined by~\cite{penrose1955generalized}:

\begin{equation}
\label{eqn:moorepenrose}
    M^{\dagger} = (M^{T}M)^{-1} M^{T}
\end{equation}
For this work, we have chosen to compare our method to FBP, SART, and the Moore-Penrose-based PI~\cite{penrose1955generalized}.
\newline

\section*{Related Work}

Over the past years, QC research has accelerated rapidly, as practical examples have come within reach. Specifically, image processing and machine learning on quantum computers have evolved to be an active area of research~\cite{caraiman2012image, biamonte2017quantum}.


In the following passage, we want to highlight work in the areas of quantum algorithm development for solving linear systems and tomographic image reconstruction. 

The reconstruction problem of tomographic data is a system of linear equations. Therefore, we first review existing approaches on quantum computers to solve them. The initial proposal to solve linear equations on a gate-based quantum computer was made by Harrow et al. in 2009 when they introduced the HHL-algorithm~\cite{harrow2009quantum}. The HHL algorithm is the basis for solving linear equations on gate-based quantum computers and is based on quantum phase estimation. Another take to solving (combinatorial) optimization problems on gate-based quantum computers is the Quantum Approximate Optimization Algorithm (QAOA)~\cite{farhi2014quantum}.

Solving linear systems of equations is also possible with QA-based systems. The first example presented was non-negative binary matrix factorization~\cite{o2018nonnegative}. Chang et al. demonstrated in 2019 that it is possible to solve polynomial equations with QA~\cite{chang2019quantum}. The approach was refined to a linear system with floating point values and floating-point division by Roger and Singleton in 2020~\cite{rogers2020floating}. Their paper utilizes the D-Wave 2000Q system to its whole extinct and shows results for matrix inversions of 3 × 3 matrices. However, they fail for matrices with high condition numbers. The first practical application to utilize QA for linear systems were Souza et al., who presented a seismic inversion problem which they solved in a least-square manner~\cite{souza2022application}.

There have been initial steps to tackle the problem of medical image reconstruction with QC. The first who proposed to perform tomographic image reconstruction with both CT, PET, and MRI were Kiani et al.~\cite{kiani2020quantum}. They proposed substituting classical Fourier transform with quantum-based Fourier transform to achieve a run time decrease~\cite{kiani2020quantum}. Moreover, Schielein et al. presented a road map towards QC-assisted CT, describing data loading, storing, image processing, and image reconstruction problems. For a long time, QA hardware needed to be more mature for realistic problems. Schielein et al. were the first to propose to solve tomographic reconstruction with QA or the QAOA~\cite{farhi2014quantum}. Following up, Jun proposed an implementation to use QA for image reconstruction in CT using sinogram-based optimization~\cite{jun2022highly}.

In this work, we want to present multiple achievements: 

\begin{itemize} 

    \item Comparison of binary and integer-based tomographic reconstruction run on actual QC hardware to classical reconstruction algorithms 

    \item Analysis of capabilities and limitations of QA regarding image size, noise, and underdetermination of the system 

    \item A framework for the creation of tomographic toy problems to accelerate quantum image reconstruction research

\end{itemize} 

\section*{Results}

\begin{figure}
     \centering
     \begin{subfigure}[b]{0.49\textwidth}
         \centering
         \includegraphics[scale = 0.25]{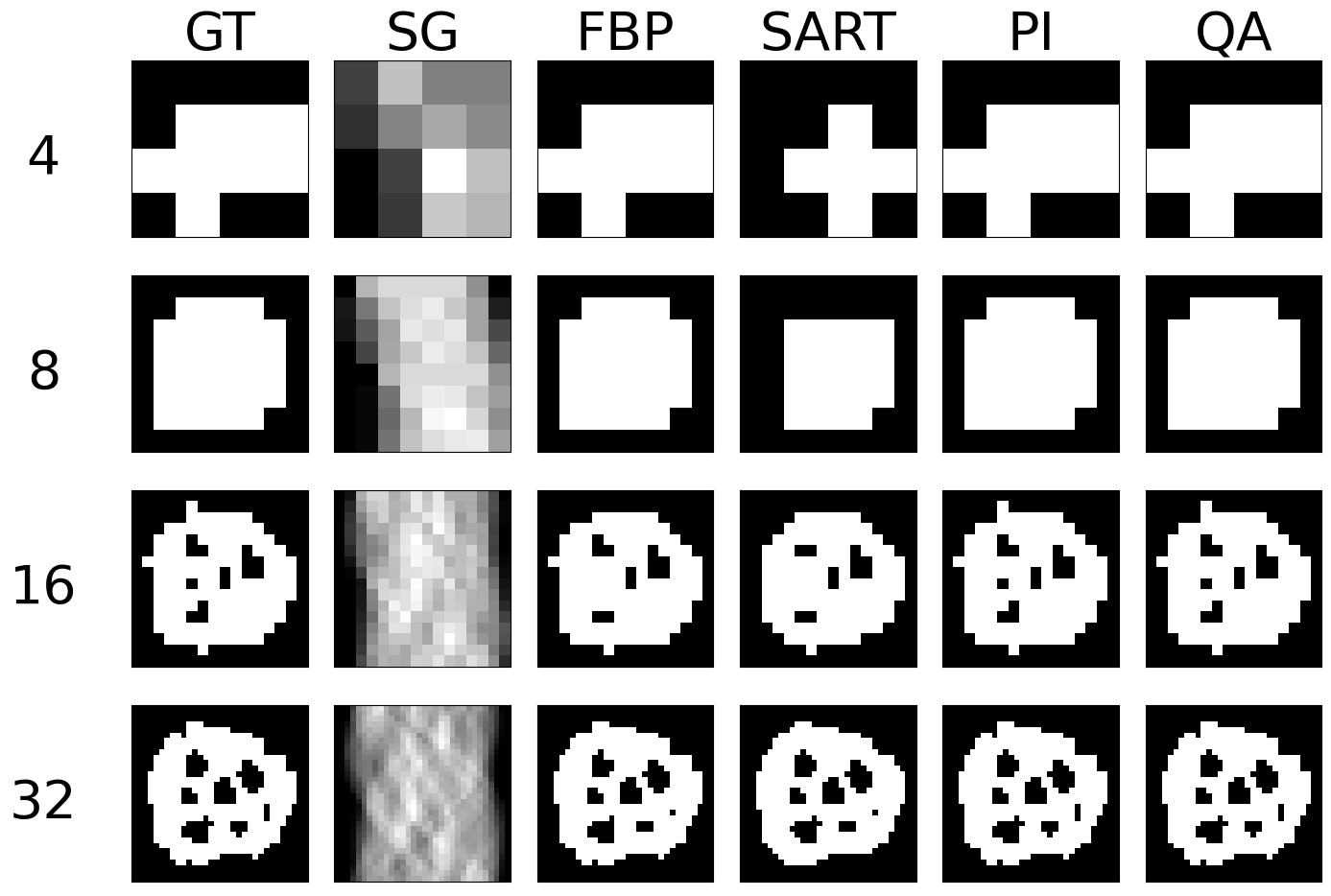}
         \label{fig:bin_foam}
     \end{subfigure}
     \hfill
     \begin{subfigure}[b]{0.49\textwidth}
         \centering
         \includegraphics[scale = 0.25]{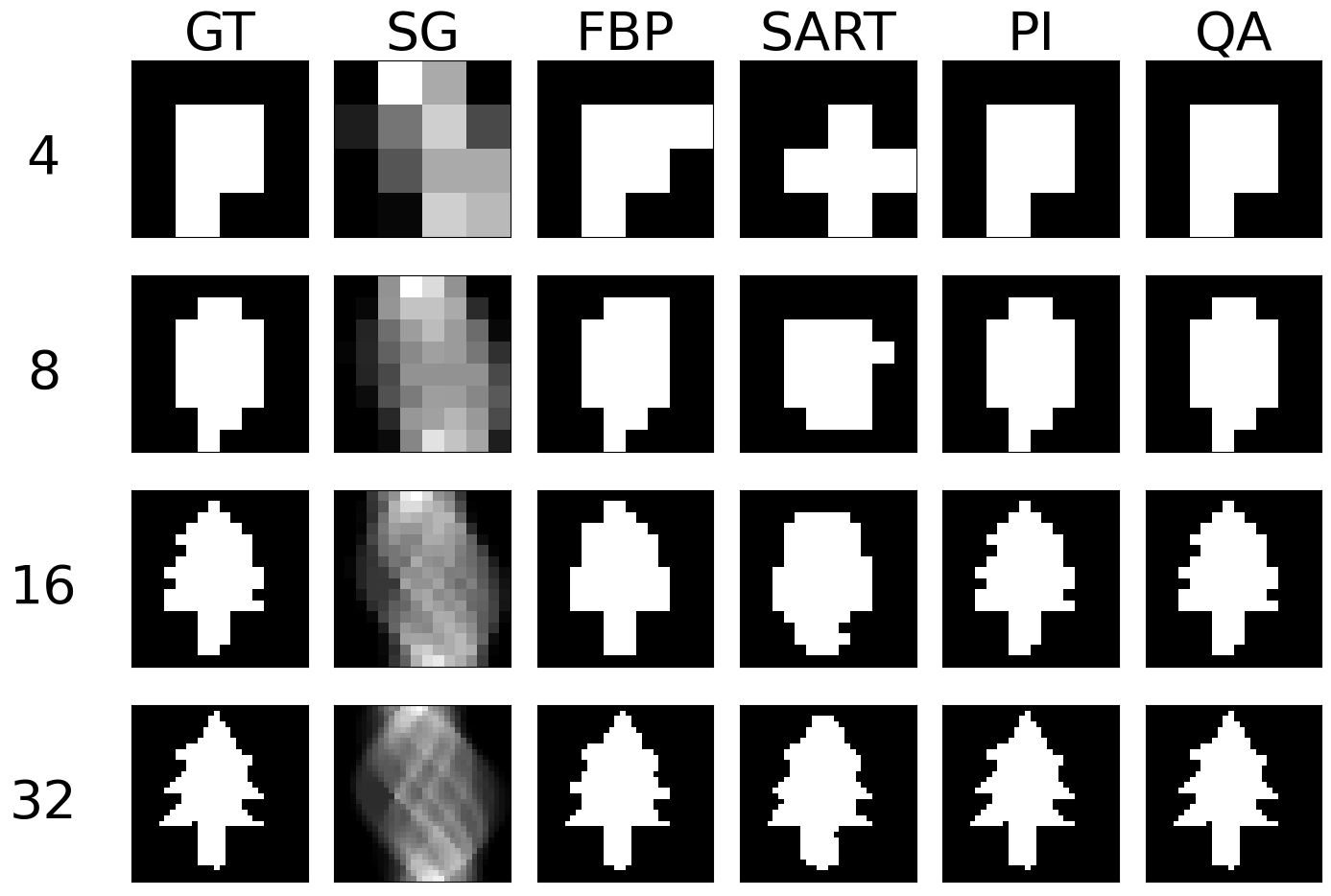}
         \label{fig:bin_tree}
     \end{subfigure}
    \caption{Binary reconstructions of sample image 'foam' (left) and 'tree' (right) for image sizes $N \times N$, where N is 4, 8, 16, and 32.}
    \label{fig:binary_mat}
\end{figure}

\begin{figure}
     \centering
     \begin{subfigure}[b]{0.49\textwidth}
         \centering
         \includegraphics[scale = 0.5]{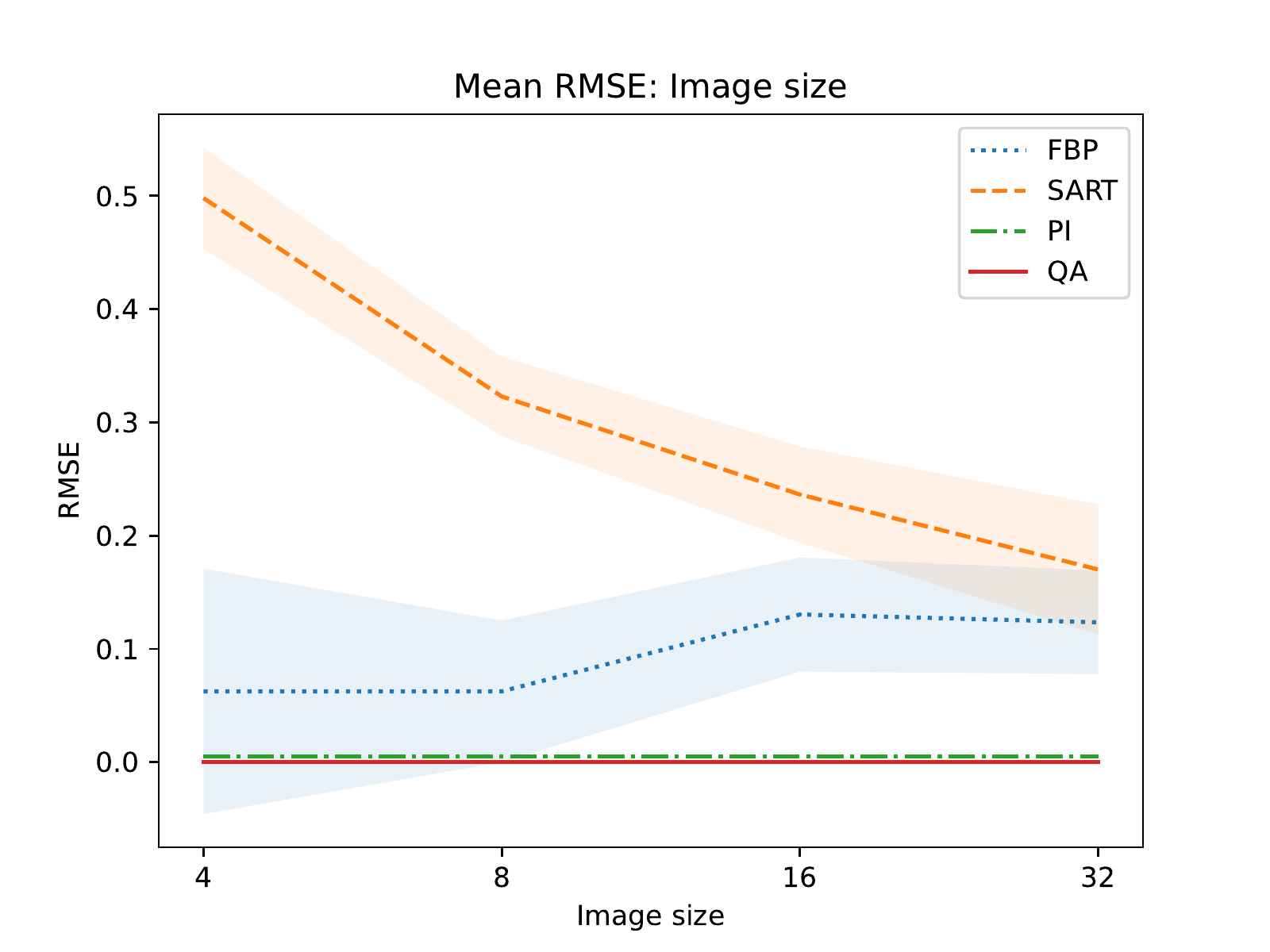}
         \label{fig:bin_mean_rmse}
     \end{subfigure}
     \hfill
     \begin{subfigure}[b]{0.49\textwidth}
         \centering
         \includegraphics[scale = 0.5]{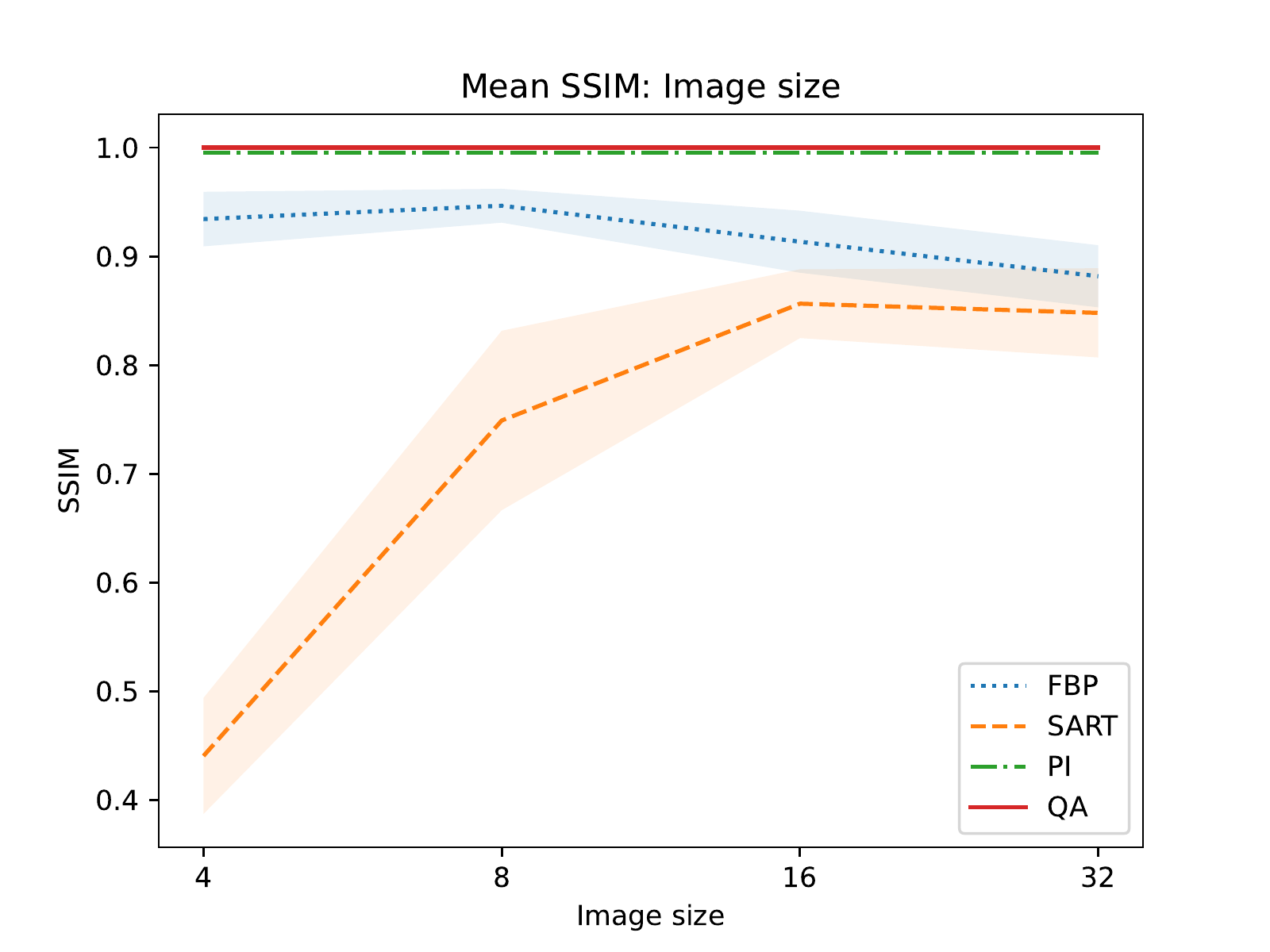}
         \label{fig:bin_mean_ssim}
     \end{subfigure}
    \caption{Mean and variance RMSE (left) and SSIM (right) evaluation of images 'foam', 'molecule', 'snowflake' and 'tree' for image sizes $N \times N$, where N is 4, 8, 16, and 32.}
    \label{fig:binary_eval}
\end{figure}

In this section, we present novel reconstruction results of our algorithm utilizing D-Wave's quantum computers and compare them to classical methods. Due to the size restrictions on the actual quantum annealer, images are reconstructed using the hybrid solvers to enable the representation of more significant problems. The size restrictions of the quantum annealer are provided in the methods section. The time limit $T$ for the hybrid optimization is fixed at $5 s$, where the annealing time is $0.015 s$. Therefore, the overhead is associated with classical computations. Every hybrid reconstruction is compared to three different classical methods: FBP, SART, and PI. We refer to our hybrid reconstruction method as QA. To compare the methods for binary tomography, we discretize the reconstruction result of the classical algorithms.
Moreover, we compare the ground truth image (GT) to the reconstructions and visualize the corresponding sinogram (SG) to the tomographic problem. We utilize SymPy for our problem formulation and solve the reconstruction problem as a classical forward and inverse problem $Mx=y$. Therefore, it applies to \textbf{any linear imaging or display system}. The system matrix for the reconstruction problem is calculated using the Radon transform. In reality, for more significant problems, the system matrix becomes infeasible to store. Nevertheless, we want to test the general performance of the hybrid solver on inverse problems regarding size, noise, and underdetermination of the linear equations.

\begin{figure}
    \centering
    \includegraphics[scale = 0.25]{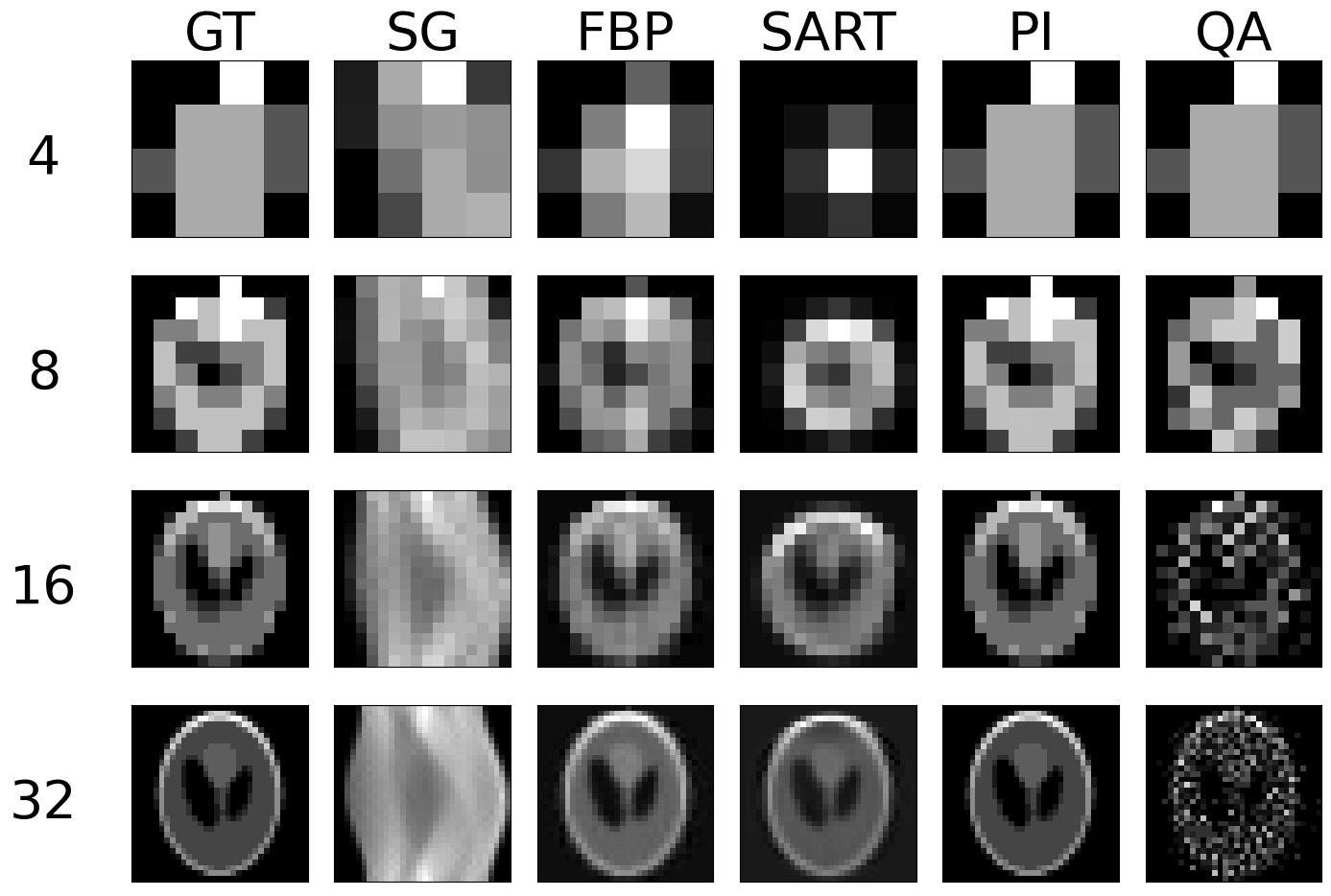}
    \caption{4-bit integer reconstructions of Shepp-Logan phantom for image sizes $N \times N$, where N is 4, 8, 16, and 32.}
    \label{fig:int_sl}
\end{figure}

\begin{figure}
     \centering
     \begin{subfigure}[b]{0.49\textwidth}
         \centering
         \includegraphics[scale = 0.5]{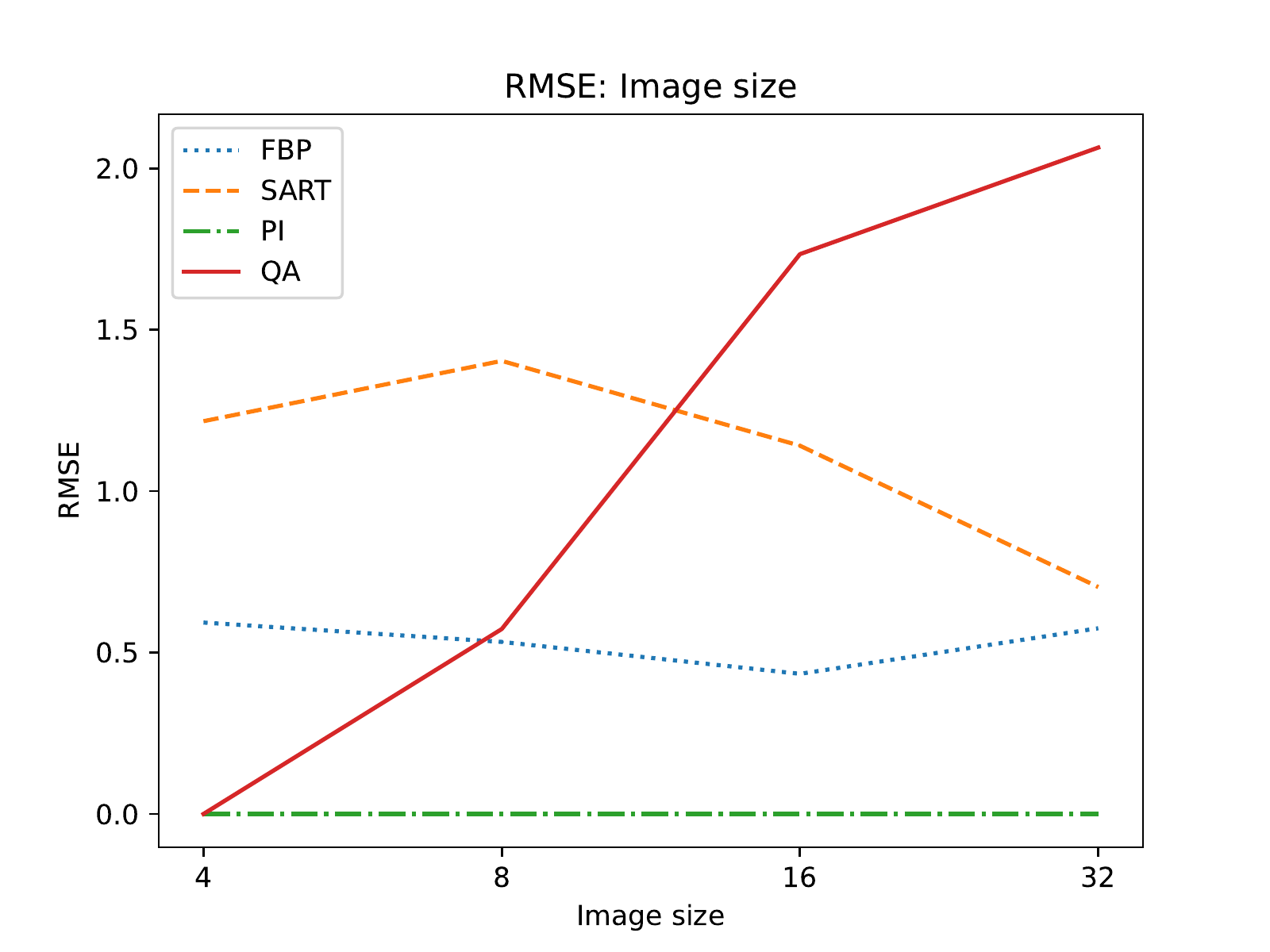}
         \label{fig:int_sl_rmse}
     \end{subfigure}
     \hfill
     \begin{subfigure}[b]{0.49\textwidth}
         \centering
          \includegraphics[scale = 0.5]{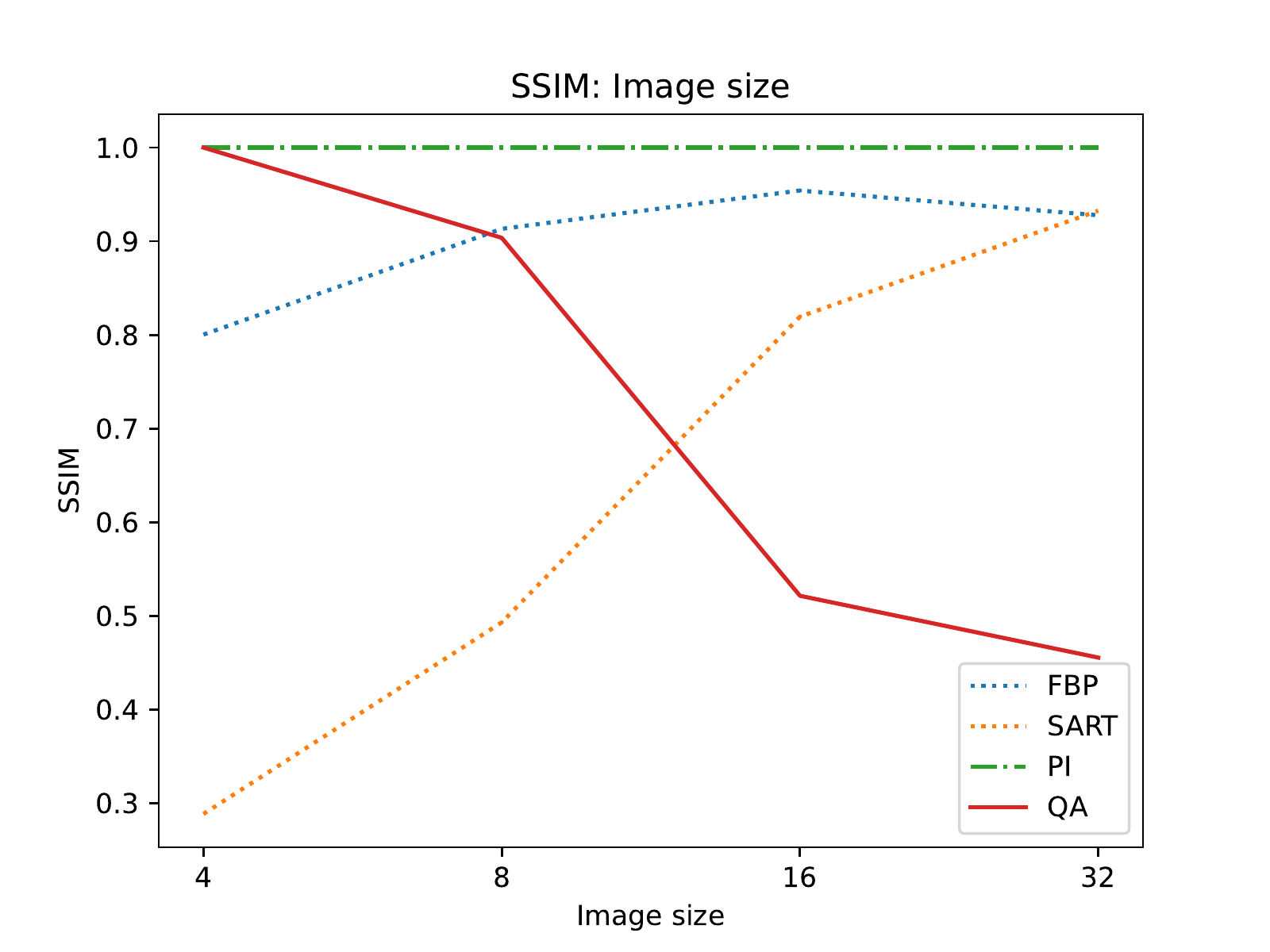}
         \label{fig:int_sl_ssim}
     \end{subfigure}
     \caption{RMSE (left) and SSIM (right) of 4-bit integer reconstructions of the Shepp-Logan phantom for image sizes $N \times N$, where N is 4, 8, 16, and 32.}
    \label{fig:int_sl_eval}
\end{figure}

\subsection*{Experimental Setup}

We have constructed a tomographic toy problem framework to test quantum computers' initial stages of image reconstruction. We set up example problems for our QA-based reconstruction by generating tomographic problems in a linear system manner. We utilize scikit-image~\cite{scikit-image} to perform Radon transforms of our ground truth images and create our system matrices. Here, the integration of the object rotated by an angle $\alpha$ defines one projection view. The number of angles equally distribute between $0^\circ$ and $180^\circ$. ET inspires our model. Therefore, we do not model the attenuation of a source light ray through the object as one would do in transmission imaging. Instead, we model the emission process of photons within the object. Here we neglect the attenuation of the photons for now.
The projection views at $0^\circ$ are taken from the top of the image. Subsequently, the angles are distributed in a clockwise direction. We have utilized scikit-image's \textit{iradon} and \textit{sart} to perform FBP and SART reconstruction algorithms. We utilize FBP with a ramp filter. For SART, we perform two iterations of the algorithm. The PI-based reconstruction uses NumPy's \textit{pinv} function to estimate a Moore-Penrose PI. 

\subsubsection*{Additive Noise}
To test problems concerning noise in the data, we establish a simple noise model to alter the projection data. Hereby, we want to imitate the statistics of a low-count ET measurement. We apply the noise in an additive manner to the ground truth image for each projection view to create independent noise realizations:

\begin{equation}
\label{eqn:addnoise}
    x_{noise} = x + n
\end{equation}
The noise is therefore defined as follows:

\begin{equation}
\label{eqn:noise}
    n_{i} = \begin{cases}
            {-1, 0, 1}, x_{i} \neq 0\\
            {0, 1}, x_{i} = 0
            \end{cases}
\end{equation}
With this noise model we want to resemble the signal dependence of Poisson noise. Poisson noise is the most prominent noise factor in projection images with very low counts.

\subsection*{Image Size Evaluation}

We test the reconstruction capabilities of the hybrid solver concerning the image size $N \times N$ of $x$. Algorithm~\ref{alg:two} describes the problem formulation for the hybrid solver. We apply our reconstruction technique to four different binary images {'foam', 'tree', 'snowflake', 'molecule'} at four different squared image sizes 4, 8, 16, and 32. We chose the images to achieve a variance in frequency and image content. In order to downsample the image, we take local means of image blocks. We take $N$ projection view with $N$ measurements for each view. Thus, we have a fully-determined system for image size $N \times N$. We show the reconstructed images and their corresponding ground truth, sinogram, and comparable classical results for the binary images 'foam' and 'tree' in Fig.~\ref{fig:binary_mat}. The examples for 'snowflake' and 'molecule' are in the appendix. Moreover, we compare the reconstruction algorithms on the four binary images measured by root mean square error (RMSE) and structural similarity index (SSIM) in Fig.~\ref{fig:binary_eval}.

We employ the hybrid solver by using integer-valued variables in a  $4$-bit range representing the numbers from $0$ to $16$. With this, we move towards a more realistic use case. We simulate the well-known Shepp-Logan phantom at the possible $4$-bit range and compare the hybrid integer reconstructions with conventional reconstruction methods in Fig.~\ref{fig:int_sl}. We perform the experiments at four different squared image sizes 4, 8, 16, and 32. In order to downsample the image, we take local means of image blocks. We take $N \times N$ projected sinogram measurements for image size $N \times N$ to achieve a fully-determined system. Further, we plot a comparison regarding RMSE and SSIM for the reconstructed image size in Fig.~\ref{fig:int_sl_eval}.

\subsection*{Noise Evaluation}

\begin{figure}
     \centering
     \begin{subfigure}[b]{0.49\textwidth}
         \centering
         \includegraphics[scale = 0.25]{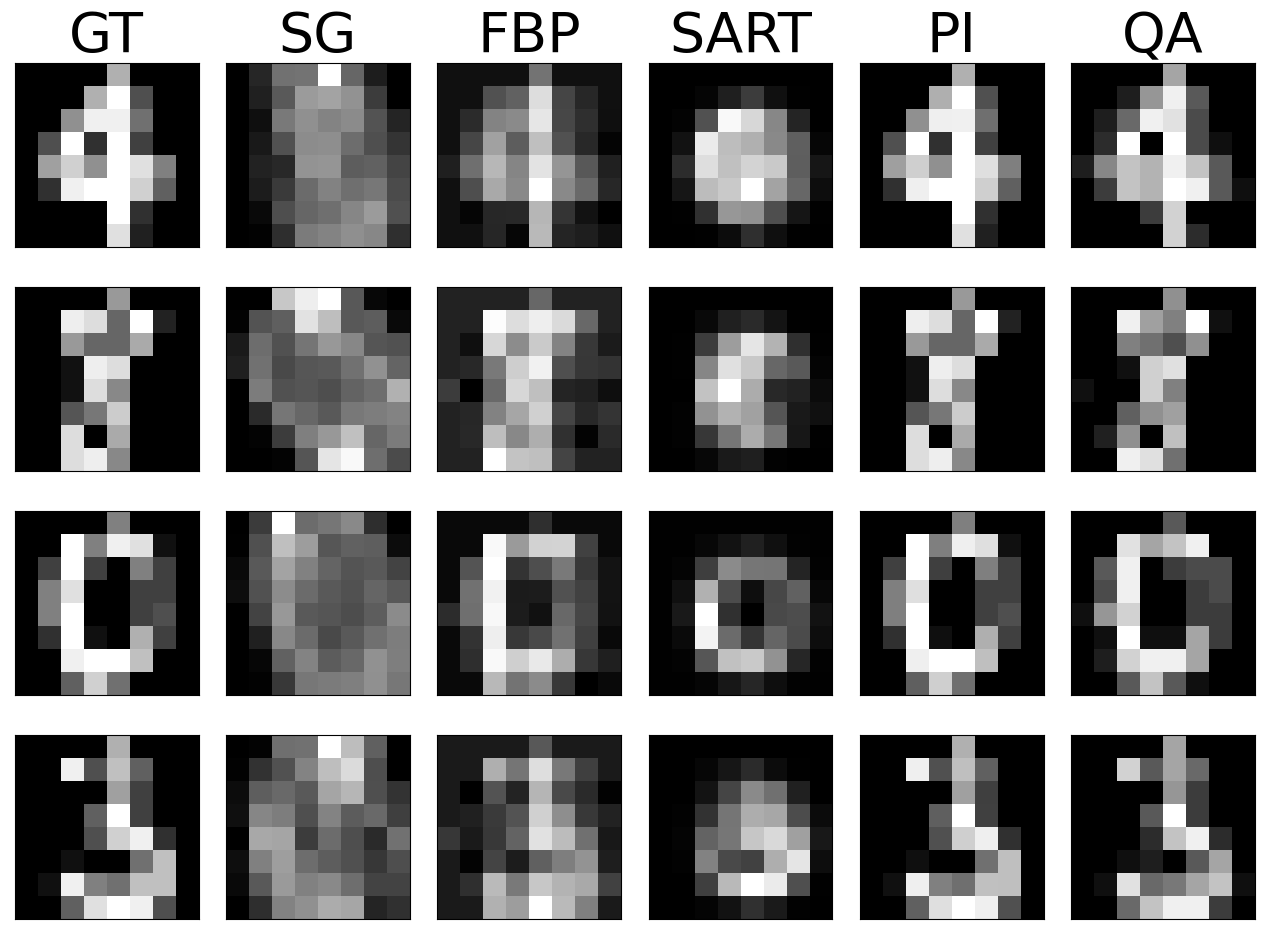}
         \label{fig:digits1}
     \end{subfigure}
     \hfill
     \begin{subfigure}[b]{0.49\textwidth}
         \includegraphics[scale = 0.25]{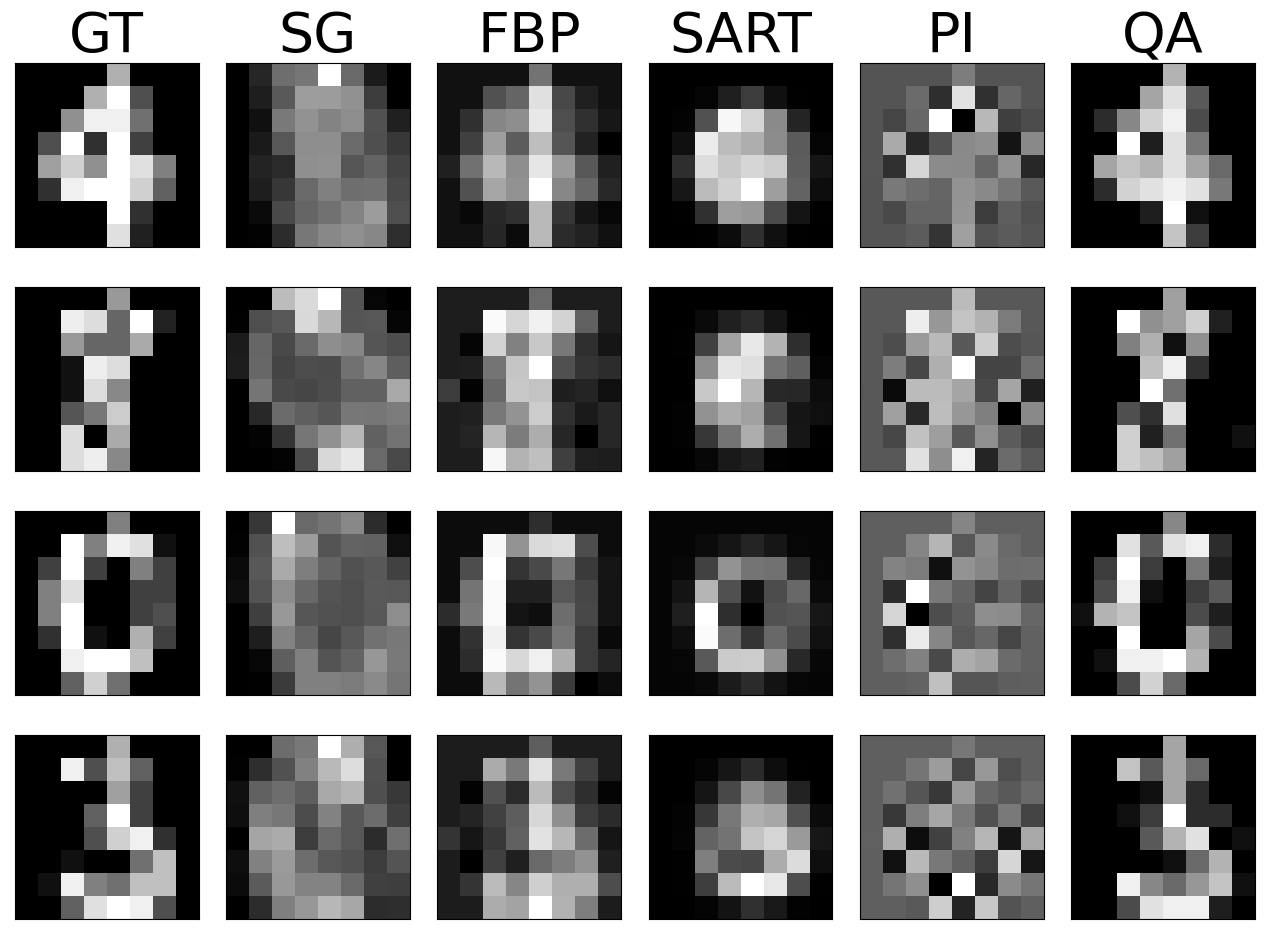}
         \label{fig:digits1_noise}
     \end{subfigure}
    \caption{4-bit integer reconstructions of four digits from the UCI digits dataset without (left) and with random noise (right).}
    \label{fig:int_digits_vis}
\end{figure}

\begin{figure}
     \centering
     \begin{subfigure}[b]{0.49\textwidth}
         \centering
         \includegraphics[scale = 0.5]{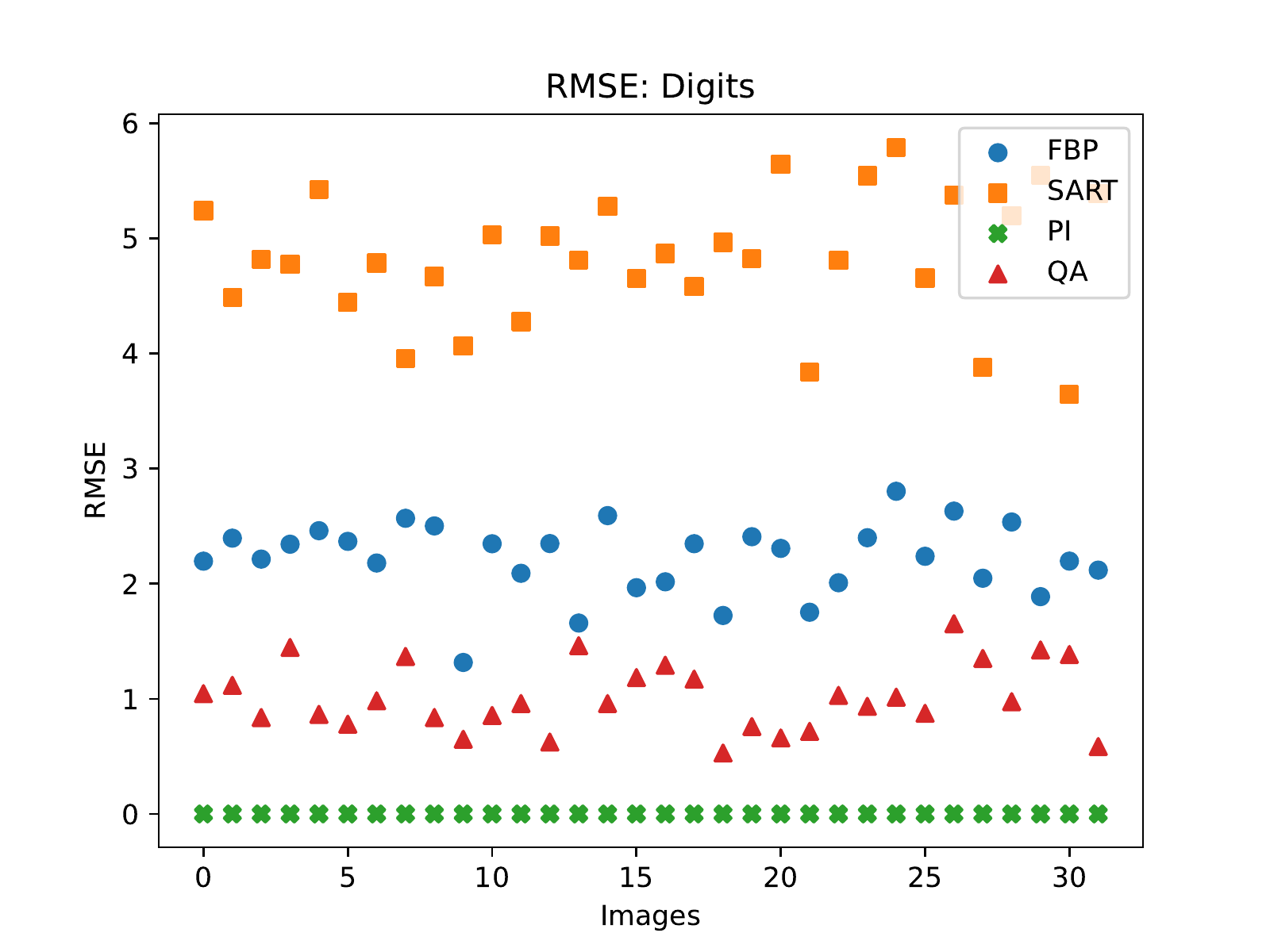}
         \label{fig:int_digits_rmse}
     \end{subfigure}
     \hfill
     \begin{subfigure}[b]{0.49\textwidth}
         \centering
          \includegraphics[scale = 0.5]{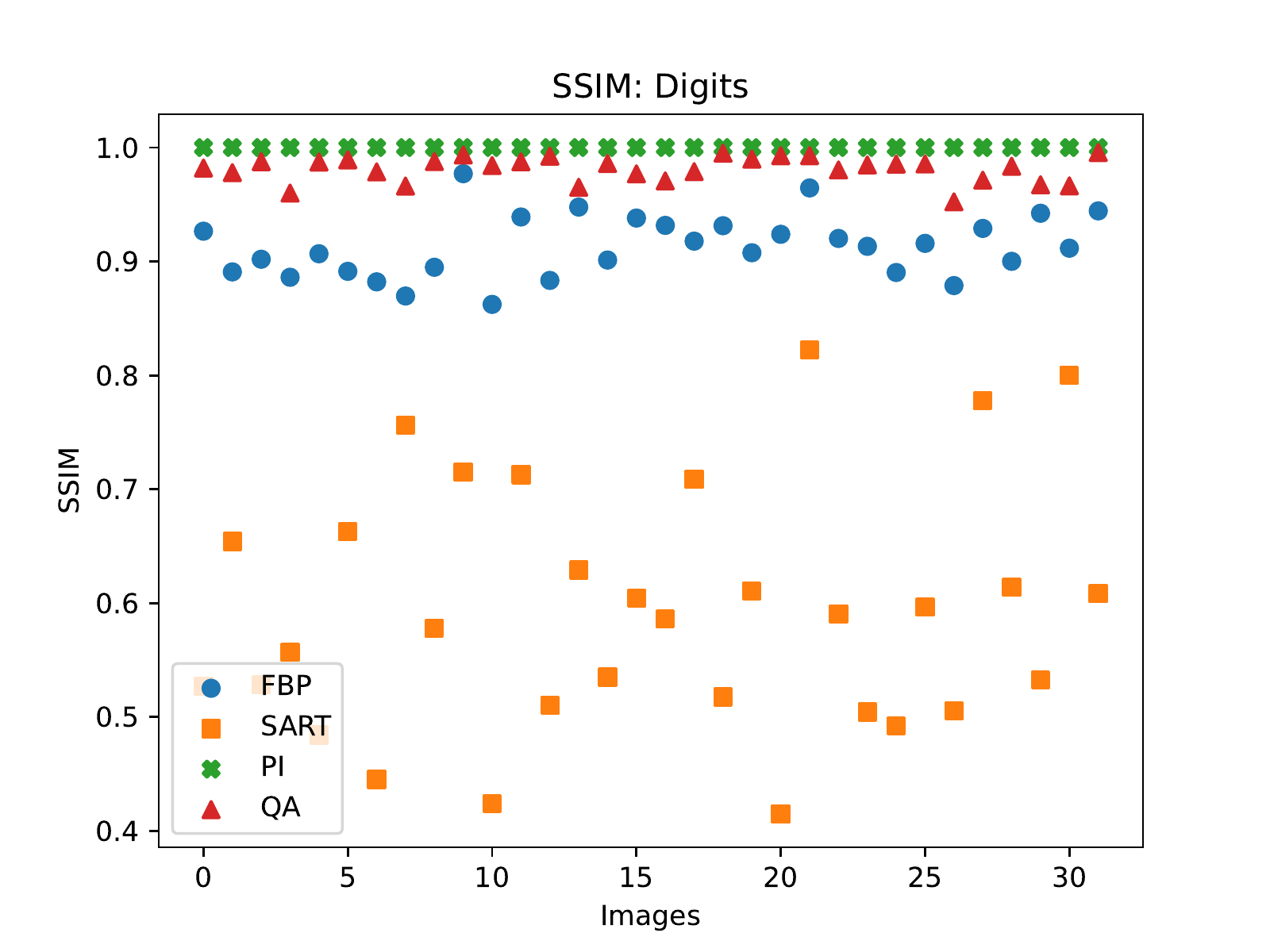}
         \label{fig:int_digits_ssim}
     \end{subfigure}
     \caption{RMSE (left) and SSIM (right) of 4-bit integer reconstructions of 32 images from the UCI digits dataset.}
    \label{fig:int_digits_eval}
\end{figure}

\begin{figure}
     \centering
     \begin{subfigure}[b]{0.49\textwidth}
         \centering
         \includegraphics[scale = 0.5]{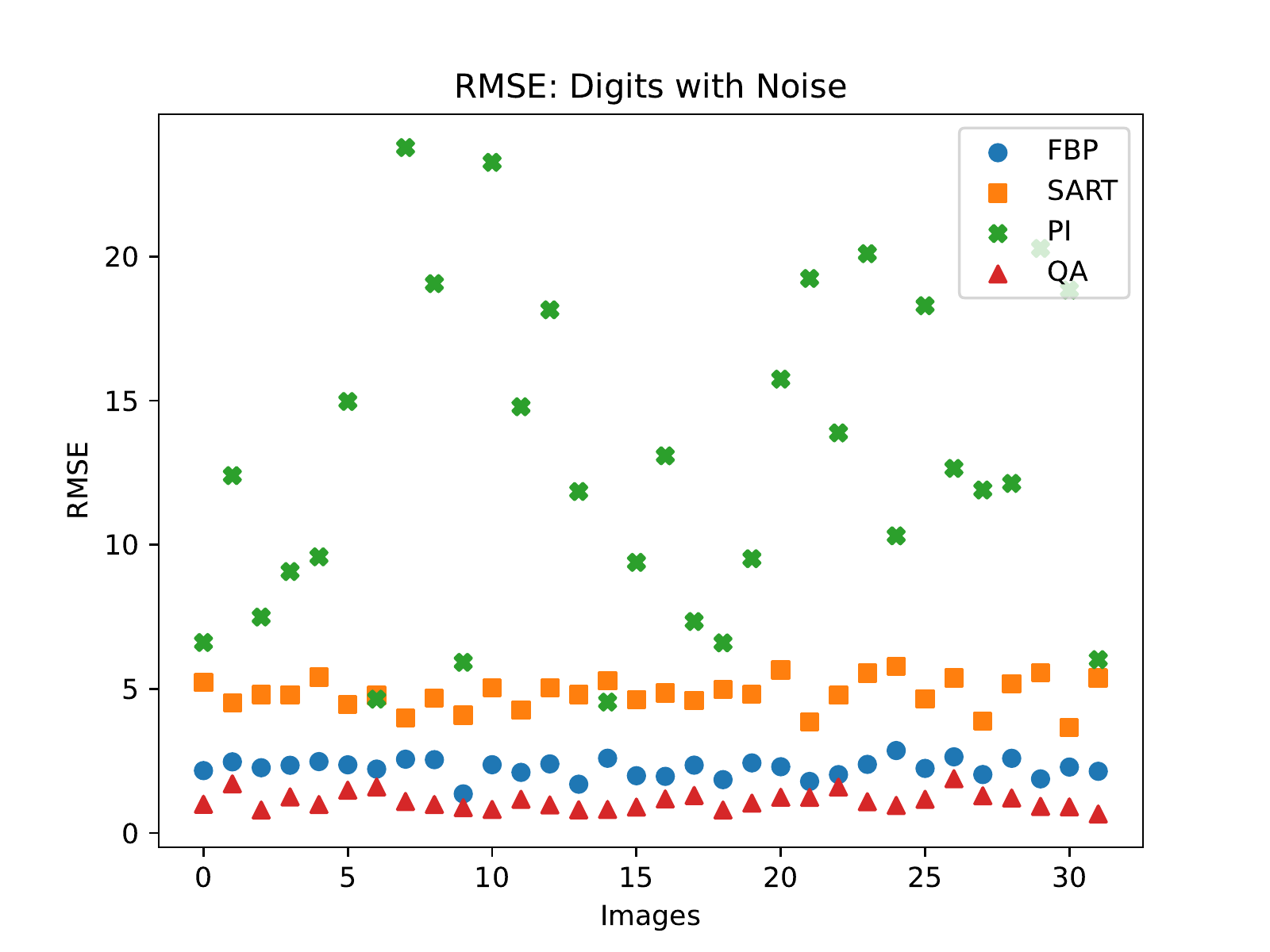}
         \label{fig:int_digits_noise_rmse}
     \end{subfigure}
     \hfill
     \begin{subfigure}[b]{0.49\textwidth}
         \centering
         \includegraphics[scale = 0.5]{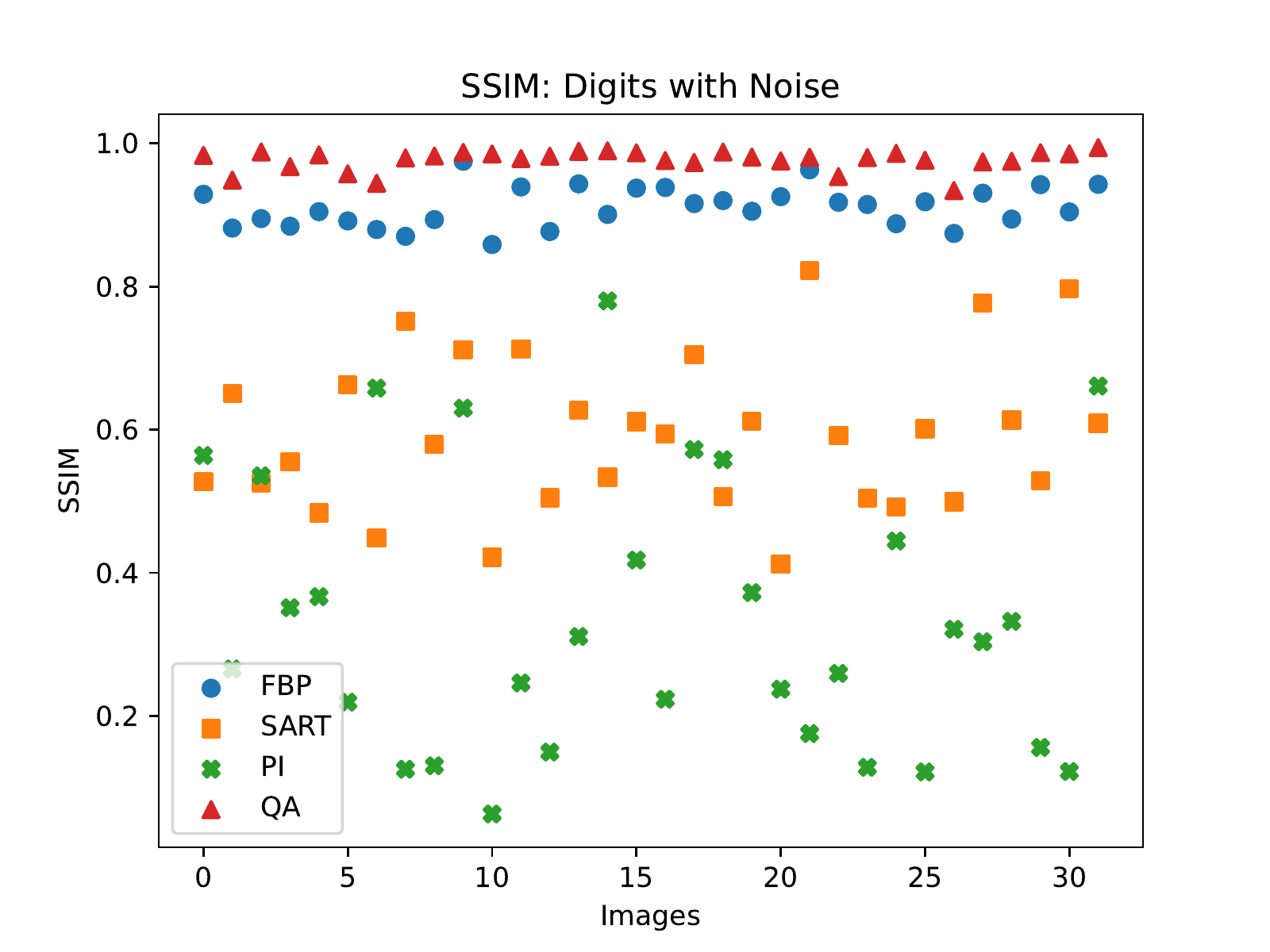}
         \label{fig:int_digits_noise_ssim}
     \end{subfigure}
     \caption{RMSE (left) and SSIM (right) of 4-bit integer reconstructions of 32 images from the UCI digits dataset with random noise.}
    \label{fig:int_digits_noise_eval}
\end{figure}

One typical problem in image reconstruction is the noise in the measured data. Especially in low-count tomography, one suffers from high photon noise. We alter image data with our noise model to imitate the high-noise level in low-count ET. We test the hybrid-based reconstruction's robustness with a simple noise alteration of the ground truth image during the acquisition. We utilize the UCI Machine Learning Repository Digits dataset~\cite{Dua:2019} for small-scale images with low bit range. The dataset consists of 5620 digits of image size $8 \times 8$ with a bit range of $[0, 16]$. We randomly chose $32$ digits and reconstructed them with and without noise. The additive noise is described in equation~\ref{eqn:addnoise}. Visual results of the reconstructed images without and with noise are shown in Fig.~\ref{fig:int_digits_vis}. Again, we show the reconstructed images and their corresponding ground truth, sinogram, and comparable classical results. We take $N \times N$ projected sinogram measurements for image size $N \times N$ to achieve a fully-determined system. The remaining reconstructed digits can be found in the appendix. Furthermore, we present a quantitative evaluation of the RMSE and SSIM for both noise-free and noisy data for each digit image in Fig.~\ref{fig:int_digits_eval} and~\ref{fig:int_digits_noise_eval}.

\subsection*{Undetermined Evaluation}

\begin{figure}
     \centering
     \begin{subfigure}[b]{0.49\textwidth}
         \centering
        \includegraphics[scale = 0.5]{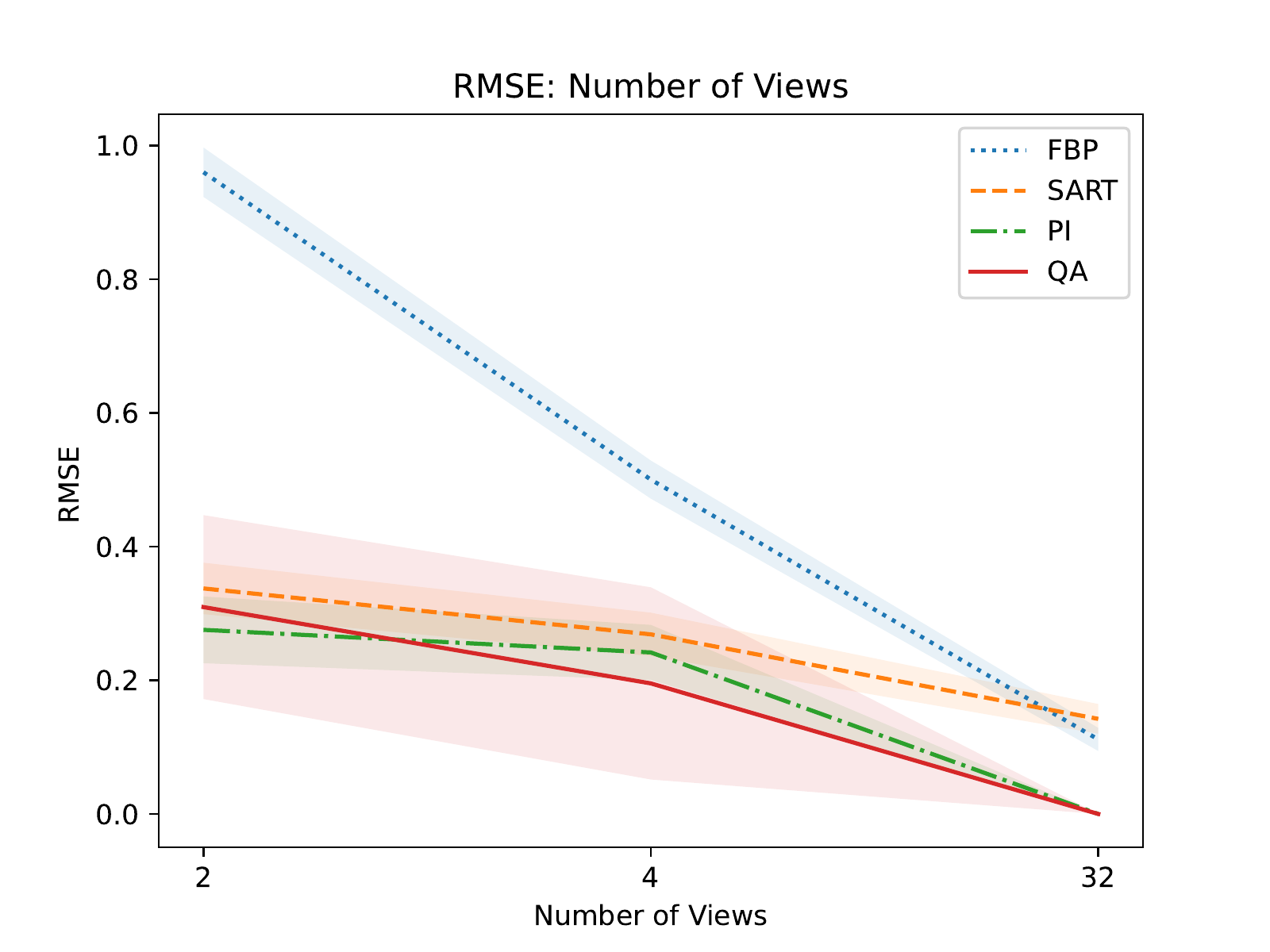}
        \caption{Mean binary reconstruction RMSE for 2, 4, and 32 views.}
        \label{fig:binary_und_rmse}
     \end{subfigure}
     \hfill
     \begin{subfigure}[b]{0.49\textwidth}
         \centering
         \includegraphics[scale = 0.5]{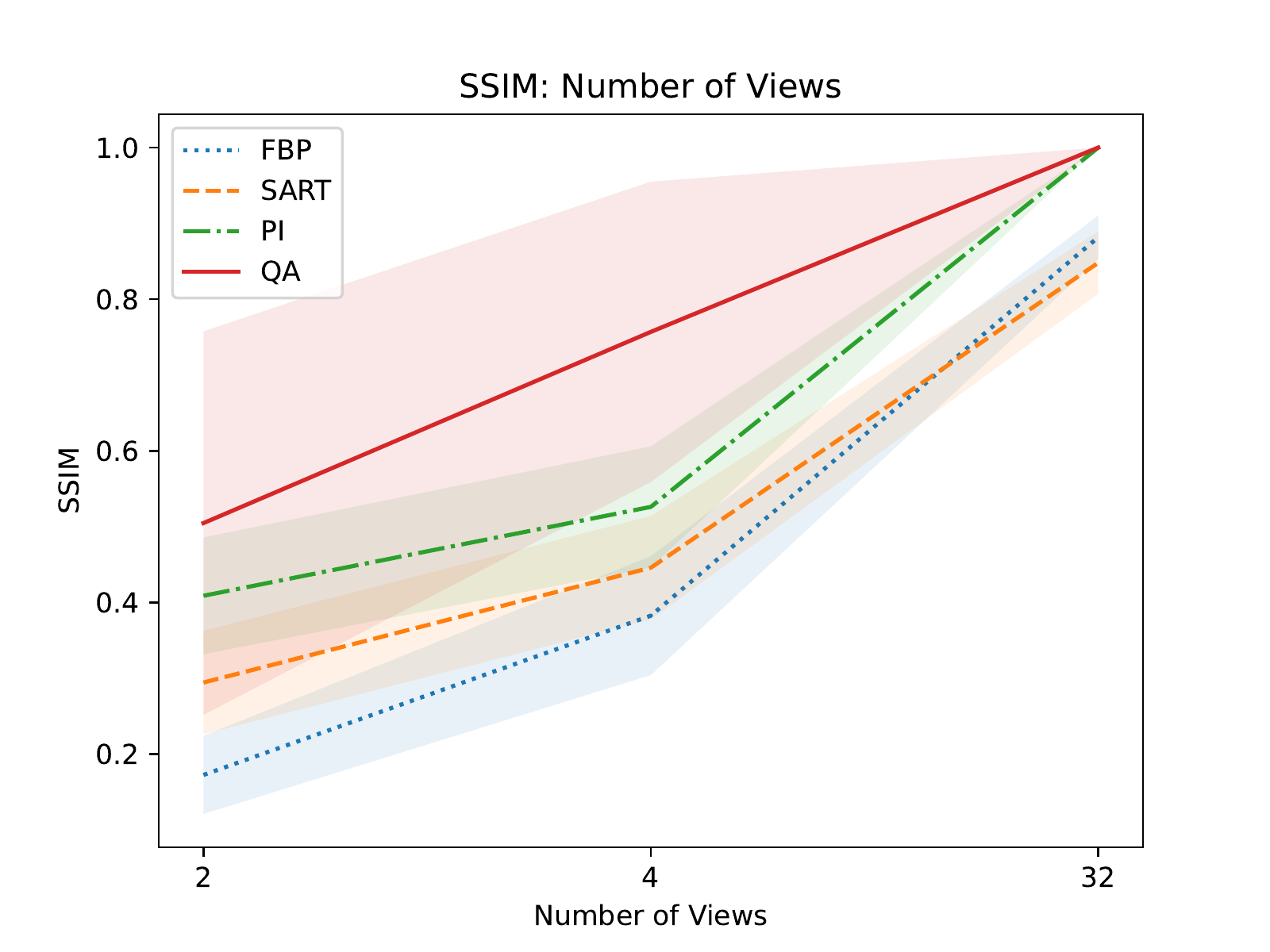}
        \caption{Mean binary reconstruction SSIM for 2, 4, and 32 views.}
        \label{fig:binary_und_ssim}
     \end{subfigure}
     \caption{RMSE (left) and SSIM (right) of binary reconstructions from 2, 4, and 32 views.}
    \label{fig:binary_und__eval}
\end{figure}

The reconstruction of binary images in a fully-determinant setting is easy for any reconstruction algorithm, as the number of combinatorial options is minimal compared to integer or floating-point-based reconstruction. The problem in binary tomography primarily results from reconstructing the objects with as few views as possible. In the past, methods have been presented to reconstruct an object with two views only~\cite{herman2012discrete}. The methods usually enforce a lot of regularization and prior knowledge of the object. Therefore, we want to present reconstruction results of binary images of size $32 \times 32$ with only $4$ and $2$ projection views acquired. The reconstructed images, their comparison algorithm results, and corresponding ground truth and sinogram are displayed for the binary image 'foam' and 'tree' in Fig.~\ref{fig:bin_und_vis}. The other examples are provided in the appendix. Moreover, we compare the reconstruction algorithms on the four binary images measured by RMSE and SSIM in Fig.~\ref{fig:binary_und__eval}.

\section*{Discussion}
Compared to previous quantum annealing-based matrix inversion, we see an improvement in linear systems with high-condition numbers. Rogers and Singleton's method of solving linear systems using a quantum annealer is restricted to small systems, $3 \times 3$, with very low condition numbers~\cite{rogers2020floating}. With the use of hybrid solvers, we can overcome this issue. Our system matrices $M$ are singular, with a condition number approaching infinity. When entirely determined, the results of binary tomographic reconstruction contest against conventional methods. We postulate that a binary image size of $32 \times 32$ is no problem for the hybrid solver. With integer-based tomographic reconstructions, we see problems approaching larger image sizes than $8 \times 8$. If we increase the solver time $T$, the reconstruction error of larger problems can be improved. Nevertheless, if we increase time $T$, the runtime of the hybrid solver cannot compete with the runtime of classical methods. 
For this reason, we performed integer-valued tomography reconstruction for the digits dataset with image size $8 \times 8$. Here we can see that the hybrid-based reconstruction can yield similar results to the conventional algorithms. However, a PI-based solution is still able to outperform the hybrid reconstruction. An exciting finding can be seen when comparing the RMSE and SSIM of noisy simulations. The hybrid-based reconstruction can outperform every conventional reconstruction technique for all 32 digits. We postulate that the energy optimization landscape is not affected much by noise during the annealing process.
Moreover, we see that the reconstruction from as few views as 2 or 4 projections can outperform standard reconstruction algorithms in binary-based images. However, the variance of the quantitative results is relatively high. The high variance can be associated with the reconstructed images, which have higher frequency content, especially within the imaged object. We postulate that the robustness to noise and the ability to reconstruct with fewer views can yield a considerable advantage, as a patient has to undergo less radiation, and less time is required to scan a patient.
We also see potential drawbacks of our method. Most importantly, the D-Wave quantum annealer and the associated hybrid solvers are no universal quantum computers. Therefore, we can only perform the QA algorithm on the hardware. In return, this means that we cannot use the ability of quantum computers to represent extensive data with significantly fewer qubits.  On the other hand, the data loading is part of the problem formulation for quantum annealer, which is a time-consuming task for gate-based QC.
Quantum annealing can only provide significant speed up and better solutions for certain problems~\cite{yan2022analytical}. The use of the hybrid solver helps improve quality solutions and problem size. However, the problems may require long runtime, which cannot compete with classical methods. In general, hybrid-based reconstruction has problems reconstructing homogeneous regions. Adding smoothness constraints to the objective could improve reconstructed images in the future. Finally, the cost of QC is relatively high at the moment, but is expected to decrease as it did for classical computers.

\begin{figure}
     \centering
     \begin{subfigure}[b]{0.49\textwidth}
         \centering
         \includegraphics[scale = 0.25]{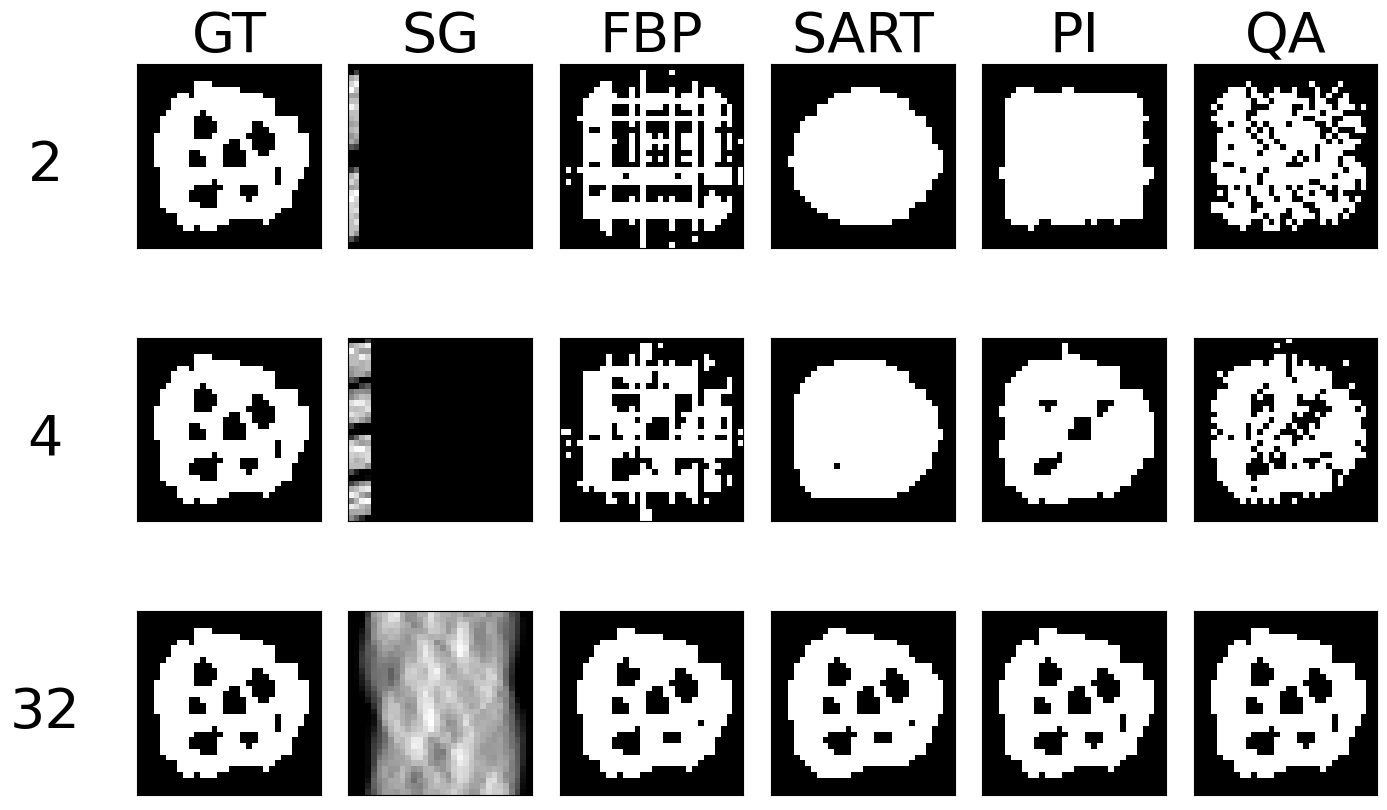}
         \label{fig:binary_und_foam}
     \end{subfigure}
     \hfill
     \begin{subfigure}[b]{0.49\textwidth}
         \includegraphics[scale = 0.25]{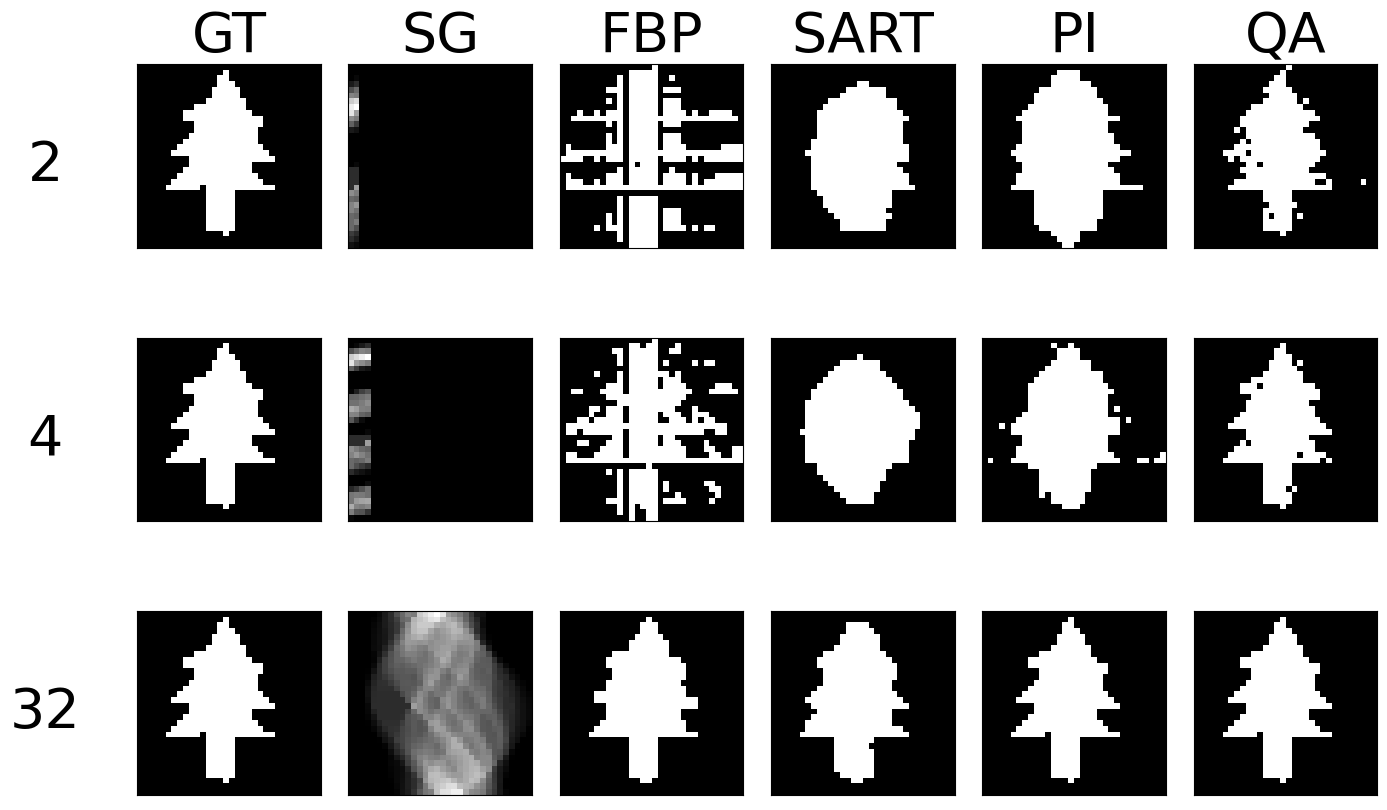}
         \label{fig:binary_und_tree}
     \end{subfigure}
    \caption{Binary reconstruction of the $32 \times 32$ image 'foam' (left) and 'tree' (right) from 2, 4, and 32 views.}
    \label{fig:bin_und_vis}
\end{figure}

\section*{Conclusion and Outlook}
We have given the reader an overview of quantum computing, especially adibatic quantum computing. More to the point, we have explained how QA works and which problems it can solve. Subsequently, we present the inverse problem of tomographic image reconstruction and describe the use-case of emission tomography and the difference to transmission tomography. We summarize previous work in the solution of linear systems and image reconstruction with quantum annealing and quantum computing and provide the fundamentals for our reconstruction method with quantum annealing and hybrid solvers. Finally, we showcase the results of binary- and integer-valued reconstruction for different matrix sizes. We also test the reconstruction concerning noise and underdetermination. Hybrid-based reconstruction can offer potential benefits in noisy linear systems and in the case of underdetermination. We also highlight the limitations due to the problem size, runtime, and explainability.

\section*{Methods}

This section will cover how we solve an inverse problem with QA. We cover the fundamentals of embedding the problem on the quantum processing unit's (QPU) topology. We elaborate on the limitations of the current topology and how hardware has to advance to run the optimization of problems at a significant scale. Further, we describe how hybrid algorithms can utilize QA to its full extent and how we have to embed the problem for the hybrid approach. Finally, we describe the open-source framework for creating noisy tomographic test problems.
\subsection*{Problem Formulation}

We recall that the reconstruction problem in tomography is an inverse problem of the form shown in equation~\ref{eqn:forwardproblem}. Therefore, the matrix inverse of M defines the solution, as seen in equation~\ref{eqn:inverseproblem}. 

  System matrices can also be non-square if an insufficient number of views is acquired, resulting in a system that is not fully determined. Therefore, we approximate the solution in a least-squares manner. In classical computing, we approximate the least-squares solution as the Moore-Penrose PI~\cite{penrose1955generalized}. For simplification, we formulate the above equation as the objective of a quadratic minimization problem, with its minimum being approximated solution of $x$:

\begin{equation}
\label{eqn:leastsquares}
\begin{aligned}
H(x) = (Mx-y)^2 = \sum_{ijk=1}^{N} M_{ki} M_{kj} x^i x^j - 2 \sum_{ij=1}^{N} y_j M_{ji} x^i + \| y \|^2 
= x^T M^T M x - x^T M^T y - y^t M x + y^T y
\end{aligned}
\end{equation} 
Previously Rogers and Singleton have described matrix inversion as a QUBO problem for floating-point precision~\cite{rogers2020floating}. Moreover, seismic inversion~\cite{souza2022application} and binary matrix factorization~\cite{o2018nonnegative} have posed similar problems.

\begin{algorithm}
\caption{Quantum annealing tomographic reconstruction}\label{alg:one}
\DontPrintSemicolon
\KwData{$\textbf{M} \in \mathbb{R}^{m \times n}$: System matrix \newline $\textbf{y} \in \mathbb{R}^{m}$: Projection vector}
\Parameter{number\_reads: Number of annealing repetitions}
\KwResult{$\textbf{x} \in \{0, 1\}^{n}$: Image vector}
$\mathbf{Linear\_Coefficients} \gets -2 \mathbf{y^{T} M }$ \\
$\mathbf{Quadratic\_Coefficients} \gets \mathbf{M^{T} M}$ \\
offset $\gets \mathbf{y^{T}y}$ \\
bqm $\gets$ BinaryQuadraticModel(\textbf{Linear\_Coefficients}, \textbf{Quadratic\_Coefficients}, offset) \\
chain\_strength $\gets$ scaled(bqm) \\
sampleset $\gets$ QuantumAnnealer.sample(bqm, number\_reads, chain\_strength) \\
\textbf{x} $\gets$ sampleset[0] \\
\textbf{return} x
\end{algorithm}

\subsection*{Quadratic Unconstrained Binary Optimization for Binary Tomography}

In the QC fundamentals, we have discussed that the Ising model is the basis and Hamiltonian for QA. In the case of binary tomography, the reconstruction problem is directly mappable to a QUBO problem. We recall that the binary tomographic model is defined like equation~\ref{eqn:forwardproblem}, where $M\in \mathbb{R}^{(m \times n)}$, $y \in \mathbb{R}^{(m)}$ and $x \in \{ 0,1 \}^n$. The linear bias values can be extracted by following equation~\ref{eqn:leastsquares}: 

\begin{equation} Q_{i,i} = -2 \sum_{j} y_j M_{ji}
\end{equation} 
Further, we can extract the coupler values as:
\begin{equation}
\label{eqn:couplers}    
Q_{i,j} = \sum_{k} M_{ki} M_{kj} 
\end{equation}
The offset should not change the minimization problem's objective but is an optional parameter for the QA-based sampler. When directly mapping a QUBO problem to the quantum annealer, one has to consider the embedding on the QPU's topology. The qubits on a Chimaera topology, as embedded on the D-Wave 2000Q, are internally connected to four other qubits and have one or two external connections to other qubits. The newer D-Wave Advantage2 system features internal connectivity of one qubit to $12$ other qubits and two to three external couplers. To map higher connectivity graphs to the QPU, one utilizes chains of qubits to represent one qubit. Fully-connected graphs are mapped to the QPU using a clique embedding~\cite{boothby2016fast}. The most extensive mappable, fully connected graph on the QPU is a graph of 65 logical qubits on the D-Wave 2000Q and 100 logical qubits on the D-Wave Advantage2. Our binary reconstruction problem, now defined as the QUBO matrix $Q$, is fully connected. More to the point, variables usually have quadratic interactions with all other variables. The limit in terms of reconstructed binary image size for the D-Wave Advantage2 is $10 \times 10$. Therefore, using a QA to reconstruct a $R$-bit integer image of size $N \times N$ without problem optimization will require $N^2 R$ fully-connected qubits. This qubit amount and connectivity will not be available soon. Gate-based quantum computers will likely offer an advantage as significantly fewer qubits are required to represent such extensive data. Fig.~\ref{fig:embedding} shows examples of the graph and corresponding embedding for image sizes of $4x4$ and $8x8$. To embed more significant problems on the D-Wave machine, we use hybrid algorithms, which are further defined. The optimization scheme is described in pseudocode~\ref{alg:one}.

\begin{algorithm}
\caption{Hybrid quantum annealing tomographic reconstruction}\label{alg:two}
\DontPrintSemicolon
\KwData{$\textbf{M} \in \mathbb{R}^{m \times n}$: System matrix \newline $\textbf{y} \in \mathbb{R}^{m}$: Projection vector}
\Parameter{time\_limit: Time limit for the optimization}
\KwResult{$\textbf{x} \in \mathbb{Z}_{\geq 0}^{n}$: Image vector}
\textbf{x} $\gets$ [$symbolic\_integer_0, ..., symbolic\_integer_n$] \\
objective $\gets (\mathbf{Mx-y})^2$\\
cqm $\gets$ ConstrainedQuadraticModel(objective) \\
cqm.add\_constraint($x \geq 0$) \\
cqm.substitute\_self\_loops() \\
sampleset $\gets$ HybridSampler.sample(cqm, time\_limit) \\
\textbf{x} $\gets$ sampleset[0] \\
\textbf{return} x
\end{algorithm}

\subsection*{Hybrid Optimization}

The intermediate step to complete quantum-assisted computation is to design hybrid algorithms to enable the embedding of significant large problems on current QC hardware. On quantum annealers, one can make use of hybrid workflows. Raymond et al.~\cite{raymond2022hybrid} introduced one type of hybrid computation. Their algorithm uses a large neighborhood local search to find subproblems in the original problem. The subproblems are then of a size that is mappable to the QPU.
Moreover, D-Wave has introduced a hybrid solver for larger problems. The new constrained quadratic models (CQM) running on D-Wave's Hybrid Sampler enable the use of integer values in the quadratic model and, therefore, drastically expand solution possibilities. The new constrained quadratic model is defined as:

\begin{equation} 
\label{eqn:cqm}
H[x] = \sum_{i=1}^{N} a_i x_i + \sum_{i=1}^{N} \sum_{i \neq j}^{N}   b_{ij} x_i x_j  + c
\end{equation}  
Here, $x_i$ is the unknown integer variable we want to optimize for, $a_i$ is the linear weight, $b_{i,j}$ is the quadratic term between $x_i$ and $x_j$ and $c$ can define possible inequality and equality constraints. In principle, the workflow is defined by the classical problem formulation and a time limit $T$~\cite{dwave2022hybrid}. The time limit $T$ is automatically calculated depending on the problem, if not specified by the user. The solvers run in parallel and utilize heuristic solvers to explore the solution space and then pass this information to a quantum module that utilizes D-Wave systems to find solutions. The QPU solutions then guide the heuristic solvers to find better quality solutions and restrict the search space. This process iteratively repeats for the specified time limit. Further, one has the possibility of introducing quadratic and linear constraints~\cite{dwave2022hybrid}. The optimization scheme is described in algorithm~\ref{alg:two}

\section*{Code availability}

The code used to simulate the tomographic data and reconstruct the tomographic images using a quantum annealer and associated hybrid solvers will be made available after acceptance. Correspondence and requests for materials should be addressed to M.A.N.

\section*{Author contributions statement}

M.A.N. and A.H.V conceived the experiments,  M.A.N conducted the experiments, M.A.N., A.H.V, M.P.R, W.G. and A.K.M analysed the results.  All authors reviewed the manuscript. 

\bibliography{references}

\begin{thebibliography}{10}
\urlstyle{rm}
\expandafter\ifx\csname url\endcsname\relax
  \def\url#1{\texttt{#1}}\fi
\expandafter\ifx\csname urlprefix\endcsname\relax\def\urlprefix{URL }\fi
\expandafter\ifx\csname doiprefix\endcsname\relax\def\doiprefix{DOI: }\fi
\providecommand{\bibinfo}[2]{#2}
\providecommand{\eprint}[2][]{\url{#2}}

\bibitem{nielsen2002quantum}
\bibinfo{author}{Nielsen, M.~A.} \& \bibinfo{author}{Chuang, I.}
\newblock \bibinfo{title}{Quantum computation and quantum information}
  (\bibinfo{year}{2002}).

\bibitem{arute2019quantum}
\bibinfo{author}{Arute, F.} \emph{et~al.}
\newblock \bibinfo{journal}{\bibinfo{title}{Quantum supremacy using a
  programmable superconducting processor}}.
\newblock {\emph{\JournalTitle{Nature}}} \textbf{\bibinfo{volume}{574}},
  \bibinfo{pages}{505--510} (\bibinfo{year}{2019}).

\bibitem{wu2021strong}
\bibinfo{author}{Wu, Y.} \emph{et~al.}
\newblock \bibinfo{journal}{\bibinfo{title}{Strong quantum computational
  advantage using a superconducting quantum processor}}.
\newblock {\emph{\JournalTitle{Physical review letters}}}
  \textbf{\bibinfo{volume}{127}}, \bibinfo{pages}{180501}
  (\bibinfo{year}{2021}).

\bibitem{aharonov2008adiabatic}
\bibinfo{author}{Aharonov, D.} \emph{et~al.}
\newblock \bibinfo{journal}{\bibinfo{title}{Adiabatic quantum computation is
  equivalent to standard quantum computation}}.
\newblock {\emph{\JournalTitle{SIAM review}}} \textbf{\bibinfo{volume}{50}},
  \bibinfo{pages}{755--787} (\bibinfo{year}{2008}).

\bibitem{raymond2022hybrid}
\bibinfo{author}{Raymond, J.} \emph{et~al.}
\newblock \bibinfo{journal}{\bibinfo{title}{Hybrid quantum annealing for
  larger-than-qpu lattice-structured problems}}.
\newblock {\emph{\JournalTitle{arXiv preprint arXiv:2202.03044}}}
  (\bibinfo{year}{2022}).

\bibitem{maier2018medical}
\bibinfo{author}{Maier, A.}, \bibinfo{author}{Steidl, S.},
  \bibinfo{author}{Christlein, V.} \& \bibinfo{author}{Hornegger, J.}
\newblock \emph{\bibinfo{title}{Medical imaging systems: An introductory
  guide}} (\bibinfo{publisher}{Springer}, \bibinfo{year}{2018}).

\bibitem{wernick2004emission}
\bibinfo{author}{Wernick, M.~N.} \& \bibinfo{author}{Aarsvold, J.~N.}
\newblock \emph{\bibinfo{title}{Emission tomography: the fundamentals of PET
  and SPECT}} (\bibinfo{publisher}{Elsevier}, \bibinfo{year}{2004}).

\bibitem{rieffel2000introduction}
\bibinfo{author}{Rieffel, E.} \& \bibinfo{author}{Polak, W.}
\newblock \bibinfo{journal}{\bibinfo{title}{An introduction to quantum
  computing for non-physicists}}.
\newblock {\emph{\JournalTitle{ACM Computing Surveys (CSUR)}}}
  \textbf{\bibinfo{volume}{32}}, \bibinfo{pages}{300--335}
  (\bibinfo{year}{2000}).

\bibitem{chow2021ibm}
\bibinfo{author}{Chow, J.}, \bibinfo{author}{Dial, O.} \&
  \bibinfo{author}{Gambetta, J.}
\newblock \bibinfo{journal}{\bibinfo{title}{Ibm quantum breaks the 100-qubit
  processor barrier}}.
\newblock {\emph{\JournalTitle{IBM Research Blog}}}  (\bibinfo{year}{2021}).

\bibitem{albash2018adiabatic}
\bibinfo{author}{Albash, T.} \& \bibinfo{author}{Lidar, D.~A.}
\newblock \bibinfo{journal}{\bibinfo{title}{Adiabatic quantum computation}}.
\newblock {\emph{\JournalTitle{Reviews of Modern Physics}}}
  \textbf{\bibinfo{volume}{90}}, \bibinfo{pages}{015002}
  (\bibinfo{year}{2018}).

\bibitem{van2001powerful}
\bibinfo{author}{Van~Dam, W.}, \bibinfo{author}{Mosca, M.} \&
  \bibinfo{author}{Vazirani, U.}
\newblock \bibinfo{title}{How powerful is adiabatic quantum computation?}
\newblock In \emph{\bibinfo{booktitle}{Proceedings 42nd IEEE symposium on
  foundations of computer science}}, \bibinfo{pages}{279--287}
  (\bibinfo{organization}{IEEE}, \bibinfo{year}{2001}).

\bibitem{farhi2001quantum}
\bibinfo{author}{Farhi, E.} \emph{et~al.}
\newblock \bibinfo{journal}{\bibinfo{title}{A quantum adiabatic evolution
  algorithm applied to random instances of an np-complete problem}}.
\newblock {\emph{\JournalTitle{Science}}} \textbf{\bibinfo{volume}{292}},
  \bibinfo{pages}{472--475} (\bibinfo{year}{2001}).

\bibitem{born1928beweis}
\bibinfo{author}{Born, M.} \& \bibinfo{author}{Fock, V.}
\newblock \bibinfo{journal}{\bibinfo{title}{Beweis des adiabatensatzes}}.
\newblock {\emph{\JournalTitle{Zeitschrift f{\"u}r Physik}}}
  \textbf{\bibinfo{volume}{51}}, \bibinfo{pages}{165--180}
  (\bibinfo{year}{1928}).

\bibitem{jansen2007bounds}
\bibinfo{author}{Jansen, S.}, \bibinfo{author}{Ruskai, M.-B.} \&
  \bibinfo{author}{Seiler, R.}
\newblock \bibinfo{journal}{\bibinfo{title}{Bounds for the adiabatic
  approximation with applications to quantum computation}}.
\newblock {\emph{\JournalTitle{Journal of Mathematical Physics}}}
  \textbf{\bibinfo{volume}{48}}, \bibinfo{pages}{102111}
  (\bibinfo{year}{2007}).

\bibitem{finnila1994quantum}
\bibinfo{author}{Finnila, A.~B.}, \bibinfo{author}{Gomez, M.},
  \bibinfo{author}{Sebenik, C.}, \bibinfo{author}{Stenson, C.} \&
  \bibinfo{author}{Doll, J.~D.}
\newblock \bibinfo{journal}{\bibinfo{title}{Quantum annealing: A new method for
  minimizing multidimensional functions}}.
\newblock {\emph{\JournalTitle{Chemical physics letters}}}
  \textbf{\bibinfo{volume}{219}}, \bibinfo{pages}{343--348}
  (\bibinfo{year}{1994}).

\bibitem{mcgeoch2014adiabatic}
\bibinfo{author}{McGeoch, C.~C.}
\newblock \bibinfo{journal}{\bibinfo{title}{Adiabatic quantum computation and
  quantum annealing: Theory and practice}}.
\newblock {\emph{\JournalTitle{Synthesis Lectures on Quantum Computing}}}
  \textbf{\bibinfo{volume}{5}}, \bibinfo{pages}{1--93} (\bibinfo{year}{2014}).

\bibitem{koshka2020comparison}
\bibinfo{author}{Koshka, Y.} \& \bibinfo{author}{Novotny, M.~A.}
\newblock \bibinfo{journal}{\bibinfo{title}{Comparison of d-wave quantum
  annealing and classical simulated annealing for local minima determination}}.
\newblock {\emph{\JournalTitle{IEEE Journal on Selected Areas in Information
  Theory}}} \textbf{\bibinfo{volume}{1}}, \bibinfo{pages}{515--525}
  (\bibinfo{year}{2020}).

\bibitem{lucas2014ising}
\bibinfo{author}{Lucas, A.}
\newblock \bibinfo{journal}{\bibinfo{title}{Ising formulations of many np
  problems}}.
\newblock {\emph{\JournalTitle{Frontiers in physics}}} \bibinfo{pages}{5}
  (\bibinfo{year}{2014}).

\bibitem{boothby2016fast}
\bibinfo{author}{Boothby, T.}, \bibinfo{author}{King, A.~D.} \&
  \bibinfo{author}{Roy, A.}
\newblock \bibinfo{journal}{\bibinfo{title}{Fast clique minor generation in
  chimera qubit connectivity graphs}}.
\newblock {\emph{\JournalTitle{Quantum Information Processing}}}
  \textbf{\bibinfo{volume}{15}}, \bibinfo{pages}{495--508}
  (\bibinfo{year}{2016}).

\bibitem{king2018observation}
\bibinfo{author}{King, A.~D.} \emph{et~al.}
\newblock \bibinfo{journal}{\bibinfo{title}{Observation of topological
  phenomena in a programmable lattice of 1,800 qubits}}.
\newblock {\emph{\JournalTitle{Nature}}} \textbf{\bibinfo{volume}{560}},
  \bibinfo{pages}{456--460} (\bibinfo{year}{2018}).

\bibitem{dwave2022ocean}
\bibinfo{author}{D-Wave}.
\newblock \bibinfo{title}{D-wave ocean software}.
\newblock \bibinfo{howpublished}{\url{https://docs.ocean.dwavesys.com}}
  (\bibinfo{year}{2022}).
\newblock \bibinfo{note}{Accessed: 2022-09-28}.

\bibitem{dwave2022leap}
\bibinfo{author}{D-Wave}.
\newblock \bibinfo{title}{D-wave leap}.
\newblock \bibinfo{howpublished}{\url{https://cloud.dwavesys.com/leap/}}
  (\bibinfo{year}{2022}).
\newblock \bibinfo{note}{Accessed: 2022-09-28}.

\bibitem{barrett2013foundations}
\bibinfo{author}{Barrett, H.~H.} \& \bibinfo{author}{Myers, K.~J.}
\newblock \emph{\bibinfo{title}{Foundations of image science}}
  (\bibinfo{publisher}{John Wiley \& Sons}, \bibinfo{year}{2013}).

\bibitem{cormack1963representation}
\bibinfo{author}{Cormack, A.~M.}
\newblock \bibinfo{journal}{\bibinfo{title}{Representation of a function by its
  line integrals, with some radiological applications}}.
\newblock {\emph{\JournalTitle{Journal of applied physics}}}
  \textbf{\bibinfo{volume}{34}}, \bibinfo{pages}{2722--2727}
  (\bibinfo{year}{1963}).

\bibitem{beylkin1987discrete}
\bibinfo{author}{Beylkin, G.}
\newblock \bibinfo{journal}{\bibinfo{title}{Discrete radon transform}}.
\newblock {\emph{\JournalTitle{IEEE transactions on acoustics, speech, and
  signal processing}}} \textbf{\bibinfo{volume}{35}}, \bibinfo{pages}{162--172}
  (\bibinfo{year}{1987}).

\bibitem{hsieh2003computed}
\bibinfo{author}{Hsieh, J.}
\newblock \emph{\bibinfo{title}{Computed tomography: principles, design,
  artifacts, and recent advances}} (\bibinfo{publisher}{SPIE press},
  \bibinfo{year}{2003}).

\bibitem{herman2012discrete}
\bibinfo{author}{Herman, G.~T.} \& \bibinfo{author}{Kuba, A.}
\newblock \emph{\bibinfo{title}{Discrete tomography: Foundations, algorithms,
  and applications}} (\bibinfo{publisher}{Springer Science \& Business Media},
  \bibinfo{year}{2012}).

\bibitem{pan2009commercial}
\bibinfo{author}{Pan, X.}, \bibinfo{author}{Sidky, E.~Y.} \&
  \bibinfo{author}{Vannier, M.}
\newblock \bibinfo{journal}{\bibinfo{title}{Why do commercial ct scanners still
  employ traditional, filtered back-projection for image reconstruction?}}
\newblock {\emph{\JournalTitle{Inverse problems}}}
  \textbf{\bibinfo{volume}{25}}, \bibinfo{pages}{123009}
  (\bibinfo{year}{2009}).

\bibitem{bruyant2002analytic}
\bibinfo{author}{Bruyant, P.~P.}
\newblock \bibinfo{journal}{\bibinfo{title}{Analytic and iterative
  reconstruction algorithms in spect}}.
\newblock {\emph{\JournalTitle{Journal of Nuclear Medicine}}}
  \textbf{\bibinfo{volume}{43}}, \bibinfo{pages}{1343--1358}
  (\bibinfo{year}{2002}).

\bibitem{lange1984reconstruction}
\bibinfo{author}{Lange, K.}, \bibinfo{author}{Carson, R.} \emph{et~al.}
\newblock \bibinfo{journal}{\bibinfo{title}{Em reconstruction algorithms for
  emission and transmission tomography}}.
\newblock {\emph{\JournalTitle{J Comput Assist Tomogr}}}
  \textbf{\bibinfo{volume}{8}}, \bibinfo{pages}{306--16}
  (\bibinfo{year}{1984}).

\bibitem{dempster1977maximum}
\bibinfo{author}{Dempster, A.~P.}, \bibinfo{author}{Laird, N.~M.} \&
  \bibinfo{author}{Rubin, D.~B.}
\newblock \bibinfo{journal}{\bibinfo{title}{Maximum likelihood from incomplete
  data via the em algorithm}}.
\newblock {\emph{\JournalTitle{Journal of the Royal Statistical Society: Series
  B (Methodological)}}} \textbf{\bibinfo{volume}{39}}, \bibinfo{pages}{1--22}
  (\bibinfo{year}{1977}).

\bibitem{hestenes1952methods}
\bibinfo{author}{Hestenes, M.~R.} \& \bibinfo{author}{Stiefel, E.}
\newblock \bibinfo{journal}{\bibinfo{title}{Methods of conjugate gradients for
  solving}}.
\newblock {\emph{\JournalTitle{Journal of research of the National Bureau of
  Standards}}} \textbf{\bibinfo{volume}{49}}, \bibinfo{pages}{409}
  (\bibinfo{year}{1952}).

\bibitem{andersen1984simultaneous}
\bibinfo{author}{Andersen, A.~H.} \& \bibinfo{author}{Kak, A.~C.}
\newblock \bibinfo{journal}{\bibinfo{title}{Simultaneous algebraic
  reconstruction technique (sart): a superior implementation of the art
  algorithm}}.
\newblock {\emph{\JournalTitle{Ultrasonic imaging}}}
  \textbf{\bibinfo{volume}{6}}, \bibinfo{pages}{81--94} (\bibinfo{year}{1984}).

\bibitem{penrose1955generalized}
\bibinfo{author}{Penrose, R.}
\newblock \bibinfo{title}{A generalized inverse for matrices}.
\newblock In \emph{\bibinfo{booktitle}{Mathematical proceedings of the
  Cambridge philosophical society}}, \bibinfo{pages}{406--413}
  (\bibinfo{organization}{Cambridge University Press}, \bibinfo{year}{1955}).

\bibitem{caraiman2012image}
\bibinfo{author}{Caraiman, S.} \& \bibinfo{author}{Manta, V.}
\newblock \bibinfo{title}{Image processing using quantum computing}.
\newblock In \emph{\bibinfo{booktitle}{2012 16th International Conference on
  System Theory, Control and Computing (ICSTCC)}}, \bibinfo{pages}{1--6}
  (\bibinfo{organization}{IEEE}, \bibinfo{year}{2012}).

\bibitem{biamonte2017quantum}
\bibinfo{author}{Biamonte, J.} \emph{et~al.}
\newblock \bibinfo{journal}{\bibinfo{title}{Quantum machine learning}}.
\newblock {\emph{\JournalTitle{Nature}}} \textbf{\bibinfo{volume}{549}},
  \bibinfo{pages}{195--202} (\bibinfo{year}{2017}).

\bibitem{harrow2009quantum}
\bibinfo{author}{Harrow, A.~W.}, \bibinfo{author}{Hassidim, A.} \&
  \bibinfo{author}{Lloyd, S.}
\newblock \bibinfo{journal}{\bibinfo{title}{Quantum algorithm for linear
  systems of equations}}.
\newblock {\emph{\JournalTitle{Physical review letters}}}
  \textbf{\bibinfo{volume}{103}}, \bibinfo{pages}{150502}
  (\bibinfo{year}{2009}).

\bibitem{farhi2014quantum}
\bibinfo{author}{Farhi, E.}, \bibinfo{author}{Goldstone, J.} \&
  \bibinfo{author}{Gutmann, S.}
\newblock \bibinfo{journal}{\bibinfo{title}{A quantum approximate optimization
  algorithm}}.
\newblock {\emph{\JournalTitle{arXiv preprint arXiv:1411.4028}}}
  (\bibinfo{year}{2014}).

\bibitem{o2018nonnegative}
\bibinfo{author}{O’Malley, D.}, \bibinfo{author}{Vesselinov, V.~V.},
  \bibinfo{author}{Alexandrov, B.~S.} \& \bibinfo{author}{Alexandrov, L.~B.}
\newblock \bibinfo{journal}{\bibinfo{title}{Nonnegative/binary matrix
  factorization with a d-wave quantum annealer}}.
\newblock {\emph{\JournalTitle{PloS one}}} \textbf{\bibinfo{volume}{13}},
  \bibinfo{pages}{e0206653} (\bibinfo{year}{2018}).

\bibitem{chang2019quantum}
\bibinfo{author}{Chang, C.~C.}, \bibinfo{author}{Gambhir, A.},
  \bibinfo{author}{Humble, T.~S.} \& \bibinfo{author}{Sota, S.}
\newblock \bibinfo{journal}{\bibinfo{title}{Quantum annealing for systems of
  polynomial equations}}.
\newblock {\emph{\JournalTitle{Scientific reports}}}
  \textbf{\bibinfo{volume}{9}}, \bibinfo{pages}{1--9} (\bibinfo{year}{2019}).

\bibitem{rogers2020floating}
\bibinfo{author}{Rogers, M.~L.} \& \bibinfo{author}{Singleton~Jr, R.~L.}
\newblock \bibinfo{journal}{\bibinfo{title}{Floating-point calculations on a
  quantum annealer: Division and matrix inversion}}.
\newblock {\emph{\JournalTitle{Frontiers in Physics}}}
  \textbf{\bibinfo{volume}{8}}, \bibinfo{pages}{265} (\bibinfo{year}{2020}).

\bibitem{souza2022application}
\bibinfo{author}{Souza, A.~M.} \emph{et~al.}
\newblock \bibinfo{journal}{\bibinfo{title}{An application of quantum annealing
  computing to seismic inversion}}.
\newblock {\emph{\JournalTitle{Frontiers in Physics}}}
  \textbf{\bibinfo{volume}{9}}, \bibinfo{pages}{748285} (\bibinfo{year}{2022}).

\bibitem{kiani2020quantum}
\bibinfo{author}{Kiani, B.~T.}, \bibinfo{author}{Villanyi, A.} \&
  \bibinfo{author}{Lloyd, S.}
\newblock \bibinfo{journal}{\bibinfo{title}{Quantum medical imaging
  algorithms}}.
\newblock {\emph{\JournalTitle{arXiv preprint arXiv:2004.02036}}}
  (\bibinfo{year}{2020}).

\bibitem{jun2022highly}
\bibinfo{author}{Jun, K.}
\newblock \bibinfo{journal}{\bibinfo{title}{Highly accurate quantum
  optimization algorithm for ct image reconstructions based on sinogram
  patterns}}.
\newblock {\emph{\JournalTitle{arXiv preprint arXiv:2207.02448}}}
  (\bibinfo{year}{2022}).

\bibitem{scikit-image}
\bibinfo{author}{van~der Walt, S.} \emph{et~al.}
\newblock \bibinfo{journal}{\bibinfo{title}{scikit-image: image processing in
  {P}ython}}.
\newblock {\emph{\JournalTitle{PeerJ}}} \textbf{\bibinfo{volume}{2}},
  \bibinfo{pages}{e453}, \doiprefix\url{10.7717/peerj.453}
  (\bibinfo{year}{2014}).

\bibitem{Dua:2019}
\bibinfo{author}{Dua, D.} \& \bibinfo{author}{Graff, C.}
\newblock \bibinfo{title}{{UCI} machine learning repository}
  (\bibinfo{year}{2017}).

\bibitem{yan2022analytical}
\bibinfo{author}{Yan, B.} \& \bibinfo{author}{Sinitsyn, N.~A.}
\newblock \bibinfo{journal}{\bibinfo{title}{Analytical solution for
  nonadiabatic quantum annealing to arbitrary ising spin hamiltonian}}.
\newblock {\emph{\JournalTitle{Nature Communications}}}
  \textbf{\bibinfo{volume}{13}}, \bibinfo{pages}{1--12} (\bibinfo{year}{2022}).

\bibitem{dwave2022hybrid}
\bibinfo{author}{D-Wave}.
\newblock \bibinfo{title}{Hybrid solvers for quadratic optimization}.
\newblock
  \bibinfo{howpublished}{\url{https://www.dwavesys.com/media/soxph512/hybrid-solvers-for-quadratic-optimization.pdf}}
  (\bibinfo{year}{2022}).
\newblock \bibinfo{note}{Accessed: 2022-10-27}.

\end{thebibliography}

\end{document}